\pdfoutput=1
\documentclass[10pt,letterpaper,journal,compsoc]{IEEEtran}
\usepackage{amsmath}
\usepackage{amsfonts}
\usepackage{amssymb}

\usepackage[usenames,dvipsnames]{color}
\usepackage{graphicx}
\usepackage{wrapfig}
\usepackage{subcaption}
\usepackage{colortbl}
\usepackage{multicol}

\usepackage[ruled,vlined]{algorithm2e}

\DontPrintSemicolon
\SetKwComment{tcp}{$\triangleright$ }{}
\SetVlineSkip{0cm}
\usepackage{hyperref}
\usepackage{breakurl}
\usepackage{url}
\makeatletter \def\url@leostyle{\@ifundefined{selectfont}{\def\UrlFont{\sf}}{\def\UrlFont{\scriptsize\ttfamily}}} \makeatother
\let\emptyset\varnothing

\begin{document}
  \title{Spatio-Temporal Linkage\\ over Location-Enhanced Services}

  \author
  {
    Fuat Bas{\i}k$^\dag$, Bu\u{g}ra~Gedik$^\dag$, \c{C}a\u{g}r{\i}~Etemo\u{g}lu$^\ddag$, Hakan~Ferhatosmano\u{g}lu$^\dag\S$\\
    $\dag$ \textsf{Department of Computer Engineering, Bilkent University, Turkey}\\
    $\S$ \textsf{Department of Computer Science, University of Warwick, UK}\\
    $\ddag$ \textsf{AveaLabs, T\"{u}rk Telekom, \.{I}stanbul, Turkey}\\
    \textsf{[fuat.basik,bgedik,hakan]@cs.bilkent.edu.tr}, \textsf{cagriozgenc.etemoglu@turktelekom.com.tr}\\
    \vspace{-0.3cm}

  }
  
    \markboth
    {IEEE TRANSACTIONS ON MOBILE COMPUTING,~VOL.~XX, NO.~XX,~XXX~2017}
    {Bas{\i}k et al.: Spatio-Temporal Linkage for Location-Enhanced Services}

  \IEEEcompsoctitleabstractindextext{\begin{abstract}
  We are witnessing an enormous growth in the volume of data generated by
  various online services. An important portion of this data contains geographic
  references, since many of these services are \emph{location-enhanced} and thus
  produce spatio-temporal records of their usage. We postulate
  that the spatio-temporal usage records belonging to the same real-world entity
  can be matched across records from different location-enhanced services. Linking
  spatio-temporal records enables data analysts and service providers to obtain information that
  they cannot derive by analyzing only one set of usage records. In this paper, we
  develop a new \emph{linkage model} that can be used to match entities from two
  sets of spatio-temporal usage records belonging to two different
  location-enhanced services. This linkage model is based on the concept of 
  $k$-$l$ \emph{diversity} --- that we developed to capture both
  spatial and temporal aspects of the linkage. To realize this linkage model in
  practice, we develop a scalable linking algorithm called
  \emph{ST-Link}, which makes use of effective spatial and temporal filtering
  mechanisms that significantly reduce the search space for matching users.
  Furthermore, \emph{ST-Link} utilizes sequential scan
  procedures to avoid random disk access and thus scales to large datasets. We
  evaluated our work with respect to accuracy and performance using several datasets. 
  Experiments show that \emph{ST-Link} is effective in practice
  for performing spatio-temporal linkage and can scale to large datasets.
    \vspace{-0.3cm}
  \end{abstract}
} 
  \maketitle

\section{Introduction}\label{sec:intro}

\noindent The size of the digital footprint left behind by users interacting
with online services is increasing at a rapid pace, due to the popularity of
location based services, social networks and related online services. An
important portion of this footprint contains spatio-temporal references and is
a fertile resource for applications for social good and business
intelligence~\cite{ref:jensen}. We refer to the services that create spatio-
temporal records of their usage as \emph{Location Enhanced Services} (LES).
For instance, Foursquare/Swarm\footnote{www.foursquare.com / www.swarmapp.com}
--- a popular social networking service, records the locations of users when
they check-in at a point-of-interest (POI) registered in the system.

We consider two varieties of LES based on a user's level of involvement in the
production of spatio-temporal usage records. Users of \emph{explicit LES}
actively participate in sharing their spatio-temporal information. Location-
based social network services, like Foursquare/Swarm, are well-known examples
of such services, where the user explicitly checks-in to a particular POI at a
particular time. On the other hand, \emph{implicit LES} produce spatio-
temporal records of usage as a byproduct of a different activity, whose focus
is not sharing the location. For instance, when a user makes a payment with
her credit card, a record is produced containing time of the payment and
location of the store. Same applies for the cell phone calls, since
originating cell tower location is known to the service provider.

We postulate that the spatial-temporal usage records belonging to the same
real-world entities can be matched across records from two different LESs.
Linking spatio-temporal records enables data scientists and service providers
to obtain information that they cannot derive by analyzing only one set of
usage records. For example, a LES provider can combine user segmentation
results derived from its own usage records with social segmentation results
derived from the publicly available Foursquare/Swarm records, assuming that
the linking of users across these two datasets can be performed effectively.
Data from multiple LES providers can be combined using common spatio-temporal
information to identify general patterns and improve urban life. While
possible, there are a number of challenges associated with performing such
linkage of entities across two spatio-temporal usage record datasets.

First, unlike in traditional record linkage~\cite{ref:survey1, ref:survey2,
ref:survey3}, where it is easier to formulate linkage based on a traditional
similarity measure defined over records (such as Minkowski distance or Jaccard
similarity), in spatio-temporal linkage similarity needs to be defined based
on time, location, and the relationship between the two. For a pair of
entities from two different datasets to be considered similar, their usage
history must contain records that are close both in space and time. Equally
importantly there must not be \emph{negative matches}, such as records that
are close in time, but far in distance. We call such negative matches,
\emph{alibi}s. To address these challenges, in this paper, we introduce a
novel \emph{linkage model} based on $k$-$l$ \emph{diversity} --- a concept we
developed to capture both spatial and temporal diversity aspects of the
linkage. Informally, a pair of entities, one from each dataset, is called
$k$-$l$ diverse if they have at least $k$ co-occurring records (both
temporally and spatially) in at least $l$ different locations. However, as we
will detail later not all co-occurring records contribute fully and equally to
the overall aggregation. Furthermore, number of alibi events of such pairs
should not exceed a predefined threshold.

Second, na\"{i}ve record linkage algorithms that compare every pair of records
take $\mathcal{O}(n^2)$ time~\cite{ref:tutorial}, where $n$ is the number of
records. However, such a computation would not scale to large dataset sizes
that are typically involved in LES. Considering that location-based social
networks get millions of updates every day, processing of hundreds of days of
data for the purpose of linkage would take impractically long amount of time.
In order to link entities in a reasonable time, the \emph{ST-Link} algorithm
we have developed uses two filtering steps before pairwise comparisons of
candidate entities are performed to compute the final linkage. Taking
advantage of the spatio-temporal structure of the data, \emph{ST-Link} first
distributes entities over coarse-grained geographical regions that we call
\emph{dominating grid cells}. Such grid cells contain most of the activities
of their corresponding entities. For two entities to link, they must have a
common dominating grid. Once this step is over, the linkage is independently
performed over each dominating grid cell. During the temporal filtering step,
\emph{ST-Link} uses a sliding window based scan to build candidate entity
pairs, while also pruning this list as alibis are encountered for the current
candidate pairs. It then performs a reverse scan to further prune the
candidate pair set by finding and applying alibis that were not known during
the forward scan. Finally, our complete linkage model is evaluated over
candidate pairs of entities that remain following the spatial and temporal
filtering steps. Pairs of entities that satisfy $k$-$l$ diversity are linked
to each other.

This paper makes the following contributions: 
\smallskip\\ 
\indent$\bullet$\textbf{Model.} We introduce a novel spatio-temporal linkage model based on
the concept of $k$-$l$ diversity for matching.
\smallskip\\
\indent$\bullet$\textbf{Algorithm}. To realize the linkage model in practice, we develop the
\emph{ST-Link} algorithm. \emph{ST-Link} applies spatial and temporal
filtering techniques to effectively prune the candidate entity pairs in order
to scale to large datasets. It also performs mostly sequential I/O to further
improve performance. 
\smallskip\\
\indent$\bullet$ \textbf{Evaluation.} We
provide an experimental study using several datasets to showcase the
effectiveness of the $k$-$l$ diversity based linkage model and the efficiency
of the \emph{ST-Link} algorithm.
\smallskip\\ 
The rest of this paper is organized as follows.
Section~\ref{sec:formalization} gives the formalization of the $k$-$l$
diversity based linkage model. Section~\ref{sec:ST-Link} explains the \emph
{ST-Link} algorithm for an effective realization of our linkage model.
Section~\ref{sec:experiments} presents the experimental evaluation.
Section~\ref{sec:related} gives the related work and
Section~\ref{sec:conclusion} concludes the paper.

\section{Spatio-temporal Linkage}
\label{sec:formalization}
\noindent
In this section, we introduce our $k$-$l$ diversity based spatio-temporal
linkage model. We first present the preliminaries, including the notation used,
and then present the detailed formalization of the linkage model.

\subsection{Notation and Preliminaries}\label{sec:prelim}
\noindent
\textbf{Datasets.} We denote the two spatio-temporal usage record 
datasets from the two LES across which the linkage is to be performed as
$\mathcal{I}$ and $\mathcal{E}$.
\medskip
\\
\noindent \textbf{Entities and events.} \emph{Entities, or users,} are real-world systems or people who
use LES. Throughout this paper, the terms user and entity will be used interchangeably. 
They are represented in the datasets with their ids, which are
different for the two LES. \emph{Events} correspond to usage records generated
by a LES as a result of users interacting with the service. For an event
$e\in\mathcal{E}$ (or $i\in\mathcal{I}$), $e.u$ (or $i.u$) represents the user
associated with the event. We use $U_\mathcal{E}$ and $U_\mathcal{I}$ to
denote the set of user ids in the datasets $\mathcal{E}$ and $\mathcal{I}$,
respectively. We have $U_\mathcal{E}=\{e.u: e\in\mathcal{E}\}$ and
$U_\mathcal{I}=\{i.u: i\in\mathcal{I}\}$.
\medskip
\\
\noindent \textbf{Location and time.} Each event in the dataset contains
location and time information. The location information is in the form of a
region, denoted as $e.r$ for event $e$. We do not use a point for location, as
for most LES the location information is in the form of a region (e.g., POI in
check-ins, cell towers in calls). The time information is a point in time,
denoted as $e.t$ for event $e$. Although an event might contain a time period
as well (e.g., call start time and duration), frequently those records contain
only the start location (e.g., originating cell tower), and thus it would be
incorrect to assume the same location for the entire duration. However, if the
time information of an event is a period, and the associated locations are
known, this event could be represented as multiple events,
each with its own location information and time point (details given in Section~
\ref{sec:timePeriod}). 
\medskip
\\
\noindent \textbf{Linkage.} Our goal is to come up with a linkage
function $\mathcal{L}$, where $\mathcal{L}(\mathcal{E},
\mathcal{I})\subseteq U_\mathcal{E} \times  U_\mathcal{I}$. Each pair in
the result,  that is $(u_1, u_2)\in \mathcal{L}(\mathcal{E},\mathcal{I})$,
represents a potential linkage. We only consider user pairs $(u_1, u_2)$ for
which there is no ambiguity in the linkage, that is $\nexists u\neq u_1 \mbox{
s.t. } (u, u_2)\in
\mathcal{L}(\mathcal{E}, \mathcal{I}) \,\wedge\, \nexists u\neq u_2 \mbox{ s.t.
} (u_1, u)\in \mathcal{L}(\mathcal{E},\mathcal{I})$.
 
\subsection{$k$-$l$ Diversity based Linkage}
\noindent
The core idea behind our linkage model is to locate pairs of users
whose events satisfy $k$-$l$ diversity. Stated informally, a pair of users is
called $k$-$l$ diverse if they have at least $k$ co-occurring events (both
temporally and spatially) in at least $l$ different locations. Furthermore,
number of alibi events of such pairs should not exceed a predefined threshold.
In what follows we provide a
number of definitions that help us formalize the proposed $k$-$l$ diversity.%
\medskip
\\
\noindent \textbf{Co-occurrence.} Two events from different datasets are
called co-occurring if they are close in space and time.
Eq.~\ref{eq:co-place} defines the $P$ relationship to capture the closeness
in space. For two records $i\in\mathcal{I}$ and $e\in\mathcal{E}$, $P$ is
defined as:
\begin{equation}
P(i,e)  \equiv (i.r \cap e.r) \neq \emptyset,
\label{eq:co-place}
\end{equation}

\noindent where $i.r$ and $e.r$ are the regions of the two events. While we
defined the closeness in terms of intersection of regions, other approaches are
possible, such as the fraction of the intersection being above a threshold:
$|i.r \cap e.r|/\text{min}(|i.r|, |e.r|) \geq \epsilon$. Our methods are equally
applicable to such measures.

Eq.~\ref{ref:co-time} defines the $T$ relationship to capture the closeness of
events in time:
\begin{equation}
T(i,e) \equiv |i.t-e.t| \le \alpha.  
\label{ref:co-time}
\end{equation}

\noindent Here, we use the $\alpha$ parameter to restrict the matching events
to be within a window of $\alpha$ time units of each other. Using
Eq.~\ref{eq:co-place} and Eq.~\ref{ref:co-time}, we define the
\emph{co-occurrence} function $C$ as:
\begin{equation}
   C(i,e) \equiv T(i,e) \wedge P(i,e)
\label{eq:co-occur}
\end{equation} 

\noindent 
\textbf{Alibi.} While a definition of similarity is necessary to link events
from two different datasets, a definition of dissimilarity is also required
to rule out pairs of users as potential matches in our linkage. Such
\emph{negative matches} enable us to rule out incorrect matches and also
reduce the space of possible matches throughout the linkage process. We refer
to these negative matches as \emph{alibi}s.

By definition alibi means \textit{``A claim or piece of evidence that one was
elsewhere when an act is alleged to have taken place''}. In this paper we use
alibi to define events from two different datasets that happened around the
same time but at different locations, such that it is not possible for a user to
move from one of these locations to the other within the duration defined by the
difference of the timestamps of the events. To formalize this, we define a
\emph{runaway} function $R$, which indicates whether locations of two events
are close enough to be from the same user based on their timestamps. We define
$R$ as follows:
\begin{equation}
R(i,e) \equiv d(i.r, e.r) \leq \lambda \cdot |i.t - e.t|
\label{ref:runaway}
\end{equation}

\noindent Here, $\lambda$ is the maximum speed constant and $d$ is a function
that gives the shortest distance between two regions. If the distance between
the regions of two events is less than or equal to the distance one can travel
at the maximum speed, then we cannot rule out linkage of users associated with
these two events. Otherwise, and more importantly, these two events form an
alibi, which proves that they cannot belong to the same user. Based on this,
we define an alibi function, denoted by $A$, as follows:
\begin{equation}
A(i,e) \equiv T(i,e) \wedge \neg P(i,e) \wedge \neg R(i,e)
\label{eq:alibi}
\end{equation}

\noindent\textbf{User linkage.} The definitions we have outlined so far are on
pairs of events, and with these definitions at hand, we can now move on to
definitions on pairs of users. Let $x\in U_{\mathcal{I}}$ and $y\in
U_{\mathcal{E}}$ be two users. We use $\mathcal{I}_x$ to denote the events of
user $x$ and $\mathcal{E}_y$ to denote the events of user $y$. In order to be
able to decide whether two users are the same person or not, we need to define
a matching between their events.

Initially, let us define the set of all co-occurring events of users $x$
and $y$, represented by the function $F$. We have:
\begin{equation} \label{eq:set}
F(x,y) = \left\{ {(i,e) \in\mathcal{I}_x \times \mathcal{E}_y: C(i,e) } \right\} 
\end{equation}

\noindent $F$ is our \emph{focus set} and contains pairs of co-occurring
events of the two users. However, in this set, some of the events may be
involved in more than one co-occurring pairs. We restrict the matching between
the events of two users by disallowing multiple co-occurring event pairs
containing the same events. Accordingly, we define $\mathcal{S}$ as the set
containing all possible subsets of $F$ satisfying this restriction. We call
each such subset an \emph{event linkage set}. Formally, we have:
\begin{equation}
\begin{split}
\mathcal{S}(x,y) &= \{S \subseteq F(x,y):\\
& \nexists \{ (i_1, e_1), (i_2, e_2)\} \subseteq S \text{ s.t. } i_1=i_2 \vee e_1=e_2\}
\end{split}
\end{equation}

\begin{figure}[!t]
\centering
\includegraphics[width=.85\linewidth]{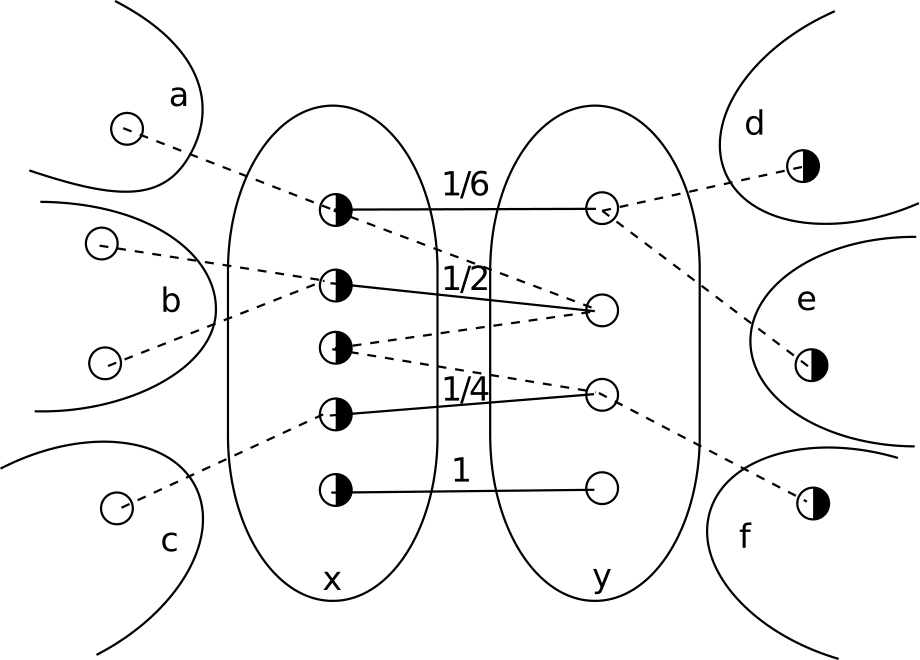}
\caption{Sample event linkage set (solid lines) for users $x$ and $y$. 
The co-occurring event pairs are shown using dashed lines. Events from a given
user are shown within circles. Users $a$, $b$, $c$, and $y$ are from one LES,
and the users $d$, $e$, $f$, and $x$ are from the other LES.}
\label{fig:match}
\end{figure}

We say that the user pair $(x, y$) satisfy $k$-$l$ diversity if there is at
least one event linkage set $S\in \mathcal{S}(x, y)$ that contains $k$
co-occurring event pairs and at least $l$ of them are at different locations.
However, each co-occurring event pair does not count as $1$, since there could
be many other co-occurring event pairs outside of $S$ or even $F$
that involve the same events. As such, we weight these co-occurring event
pairs (detailed below).
Figure~\ref{fig:match} shows a sample event linkage set with weights
for the co-occurring event pairs.
\medskip

\noindent\textbf{$k$ co-occurring event pairs.}
Let $S$ be an event linkage set in $\mathcal{S}(x,y)$ and let $\mathcal{C}$ be
a function that determines whether the co-occurring event pairs in $S$
satisfy the co-occurrence condition of $k$-$l$ diversity. We have:
\begin{equation} \label{eq:cooccur}
\mathcal{C}(S) \equiv \sum_{(i,e) \in S} w(i,e) \geq k 
\end{equation}

The weight of a co-occurring event pair is defined as:
\begin{equation}
\label{eq:weight}
\begin{split}
 w(i,e) =  &|\{i_1.u: C(i_1,e) \wedge i_1\in \mathcal{I}\}|^{-1} \cdot \\ 
           &|\{e_1.u: C(i,e_1) \wedge e_1\in \mathcal{E}\}|^{-1}
\end{split}
\end{equation}

Here, given a co-occurring event pair between two users, we check how many
possible users' events could be matched to the these events. For instance,
in the figure, consider the solid line at the top with the weight $1/6$.
The event on its left could be matched to events of $2$ different users,
and the event on its right could be matched to events of $3$ different users.
To compute the weight of a co-occurring pair, we multiply the inverse of these
user counts, assuming the possibility of matching from both sides are independent.
As such, in the figure, we get $1/2\cdot 1/3=1/6$.
\medskip
\\
\noindent\textbf{$l$ diverse event pairs.} 
\noindent For $S\in \mathcal{S}$$(x,y)$ to be $l$-diverse, there needs to be
at least $l$ unique locations for the co-occurring event pairs in it. However,
for a location to be counted towards these $l$ locations, the weights of the
co-occurring event pairs for that location must be at least 1. Let
$\mathcal{D}$ denote the function that determines whether the co-occurring
event pairs in $S$ satisfy the diversity condition of $k$-$l$ diversity. We
have:
\begin{equation}
\label{eq:diversity}
\mathcal{D}(S) \equiv | \{ p\in \mathcal{P} : \sum_{\stackrel{(i,e) \in S \text{ s.t. }}
{p\,\cap\, i.r \,\cap\, e.r \neq \emptyset }} w(i,e) \geq 1 \} | \geq l
\end{equation}

Here, one subtle issue is defining a unique location. In
Eq.~\ref{eq:diversity} we use $\mathcal{P}$ as the set of all unique locations. This
could simply be a grid-based division of the space. In our experiments, we use
the regions of the Voronoi diagram formed by cell towers as our set of unique
locations.

\begin{figure*}[!t]
  \centering
  \includegraphics[width=\linewidth]{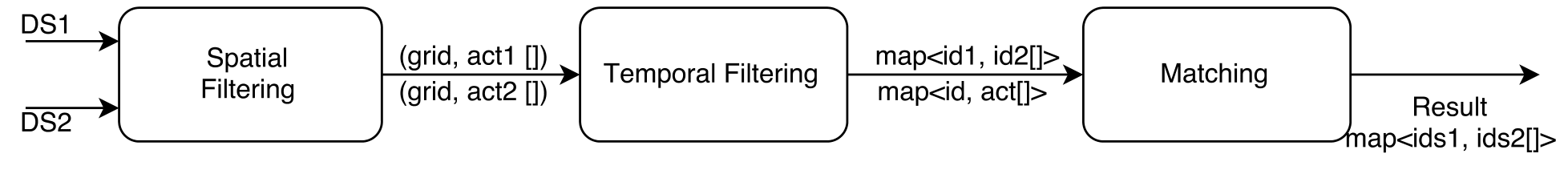}
  \caption{Data processing pipeline of ST-Link.}\label{fig:data-pipeline}
\end{figure*}

Before we can formally state the $k$-$l$ diversity based linkage, we have to
define the alibi relation for user pairs. Let $\mathcal{A}$ denote a function
that determines whether there are more than $a$ alibi events for a given pair of
users. Intuitively, having a single alibi is enough to decide that user $x$ and
$y$ are not the same person, but when there is inaccurate information, 
disregarding candidate pairs with a single alibi event might lead to false negatives.
We have:
\begin{equation}\label{eq:noAlibi}
\mathcal{A}(x,y) \equiv |{i,e} \in \mathcal{I}_x \times
\mathcal{E}_y, s.t.  A(i,e) | \leq a
\end{equation}
With these definitions at hand, we can define the spatio-temporal linkage
function $\mathcal{M}$ that determines whether users $x$ and $y$ satisfy
$k$-$l$ diversity as follows:
\begin{equation}
\label{eq:k-l-match}
\mathcal{M}(x,y) \equiv \neg \mathcal{A}(x,y) \,\wedge\, S\in \mathcal{S}(x,
y) \;s.t.\; (\mathcal{C}(S) \wedge \mathcal{D}(S))
\end{equation}

Finally, the linkage function $\mathcal{L}$ from the original problem
formulation from Section~\ref{sec:prelim} can be defined to contain only
matching pairs of users based on $\mathcal{M}$, such that there is no
ambiguity. Formally:
\begin{equation}
\label{eq:final}
\begin{split}
\mathcal{L}(\mathcal{E},\mathcal{I}) &= \{(x, y)\in \mathcal{E}\times\mathcal{I}:  
\mathcal{M}(x,y) \,\wedge\,\\
&\nexists z\neq x \text{ s.t. } \mathcal{M}(z,y) \,\wedge\,
\nexists z\neq y \text{ s.t. } \mathcal{M}(x,z)\}
\end{split}
\end{equation}

\subsection{Example Scenario}

Consider three colleagues Alice, Bob, and Carl who are working in the same
office. Assume that they all use two \emph{LES}s: \emph{les1} and \emph{les2}.
Both services generate spatio-temporal records only when they are used. The
service provider would like to link the profiles of users common in both
services. However, Bob uses the services only when he is at the office. On the
other hand, Alice and Carl use the services frequently while at work, at home,
and during vacations. Let us also assume that Alice and Carl live on the same
block, but they take vacations at different locations.

When records of Alice from \emph{les1} are processed against records of Carl
from \emph{les2}, we will encounter \emph{co-occurrences} with some amount of
diversity, as they will have matching events from work and home locations. However,
we will encounter alibi events during vacation time. In this case,
\emph{alibi} checks will help us rule out the match.

When records of Alice from \emph{les1} are processed against records of Bob
from \emph{les2}, the number of \emph{co-occurrence}s will be high, as they
are working in the same office. Yet, \emph{diversity} will be low, as Bob does
not use the services outside of the office. This also means we will not
encounter any \emph{alibi} events with Alice. In this case, diversity will
help us rule out the match.

In contrast to these cases, when Alice's own usage records from \emph{les1}
and \emph{les2} are processed, the resulting \emph{co-occurrences} will
contain high diversity since Alice uses the services at work, home, and during
vacations, and will contain no \emph{alibi}s.

In this example scenario, high number of \emph{co-occurrences} helped us 
distinguish between mere coincidences and potential candidate pairs. The
\emph{alibi} definition helped us to eliminate a false link between Alice and
Carl. Finally, \emph{diversity} helped us to eliminate a false link between
Alice and Bob, even in the absence of alibi events.


\section{ST-Link}
\label{sec:ST-Link}

\noindent In this section, we describe how the \emph{ST-Link} algorithm implements 
$k$-$l$ diversity based spatio-temporal linkage in practice. At a high-level,
\emph{ST-Link} algorithm performs filtering to reduce the space of possible entity
matches, before it performs a more costly pairwise comparison of entities
according to the formalization given in Section~\ref{sec:formalization}. The
filtering phase is divided into two steps: \emph{temporal filtering} and 
\emph{spatial filtering}. The final phase of pairwise comparisons is called
\emph{linkage}.

\subsection{Overview}\label{sub:overview}
\noindent Na\"{i}ve algorithms for linkage repeatedly compare pairwise
records, and thus take $\mathcal{O}(n^2)$~\cite{ref:tutorial} time, where $n$
is the number of records. Such algorithms do not scale to large datasets. To
address this issue, many linkage algorithms introduce some form of pruning,
typically based on blocking~\cite{ref:blocking1, ref:blocking2, ref:payger} or indexing~\cite{ref:index1, ref:index2}. Identifying
the candidate user pairs on which the full linkage algorithm is to be run can
significantly reduce the complexity of the end-to-end algorithm. Accordingly,
\emph{ST-Link} algorithm incorporates pruning strategies, which are integrated into
the spatial filtering and temporal filtering steps.

Figure~\ref{fig:data-pipeline} shows the pipelined processing of the
\emph{ST-Link} algorithm. Given two sources of data for location-enhanced services (DS1
and DS2 in the figure), spatial filtering step maps users to coarse-grained
geographical grid cells that we call \emph{dominating grid cells}. Such cells
contain most activities of the corresponding entities. Once this step is over, the remaining
steps are independently performed for each grid.

The temporal filtering step slides a window over the time ordered events to
build a set of candidate entity pairs. During this processing, it also prunes as
many entity pairs as possible based on alibi events. As we will detail later in
this section, a reverse window based scan is also performed to make sure that
all relevant alibis are taken into account.

Following the spatial and temporal filtering steps, the complete linkage is
performed over the set of candidate entity pairs. With a significantly reduced
entity pair set, the number of compared events decreases significantly as well.
Given two datasets $\mathcal{I}$ and $\mathcal{E}$, the linkage step
calculates $\mathcal{L}(\mathcal{E},\mathcal{I})$ as given in
Eq.~\ref{eq:final} without considering all possible entity pairs.

\subsection{Spatial Filtering}

\noindent By their nature, spatio-temporal data are distributed
geographically. Spatial filtering step takes advantage of this, by
partitioning the geographical region of the datasets into coarse-grained grid
cells using a modified version of quad trees~\cite{ref:quadTree}.  Each entity
is assigned to one (an in rare cases to a few) of the grid cells, which
becomes that entity's dominating grid. The dominating grid of an entity is the
cell that contains the most events from the entity. Entities that do not share
their dominating grid cells are not considered for linkage. The intuition
behind this filtering step is that, if entity $x$ from dataset $\mathcal{E}$
and entity $y$ from dataset $\mathcal{I}$ have differing dominating grids,
then they cannot be the same entity.

\subsubsection{Coarse Partitioning} \noindent For quad-tree generation in the
\emph{ST-Link} algorithm, we continue splitting the space until the grid cells
size hits a given minimum. For our experiments, we make sure that the area of
the grid cells is at least 100 km squares. For users, the grid cells should be
big enough to cover a typical user's mobility range around his home and work
location. If the minimum grid cell size is too small, then the spatial
filtering can incorrectly eliminate potential matches, as the dominating grids
from different datasets may end up being different. A concrete example is a
user that checks in to coffee shops and restaurants around his work location,
but uses a location-based match-making application only when he is at home.

\begin{figure}[h!t]
  \centering
  \begin{subfigure}{0.475\linewidth}
    \centering
    \includegraphics[height=0.6\linewidth]{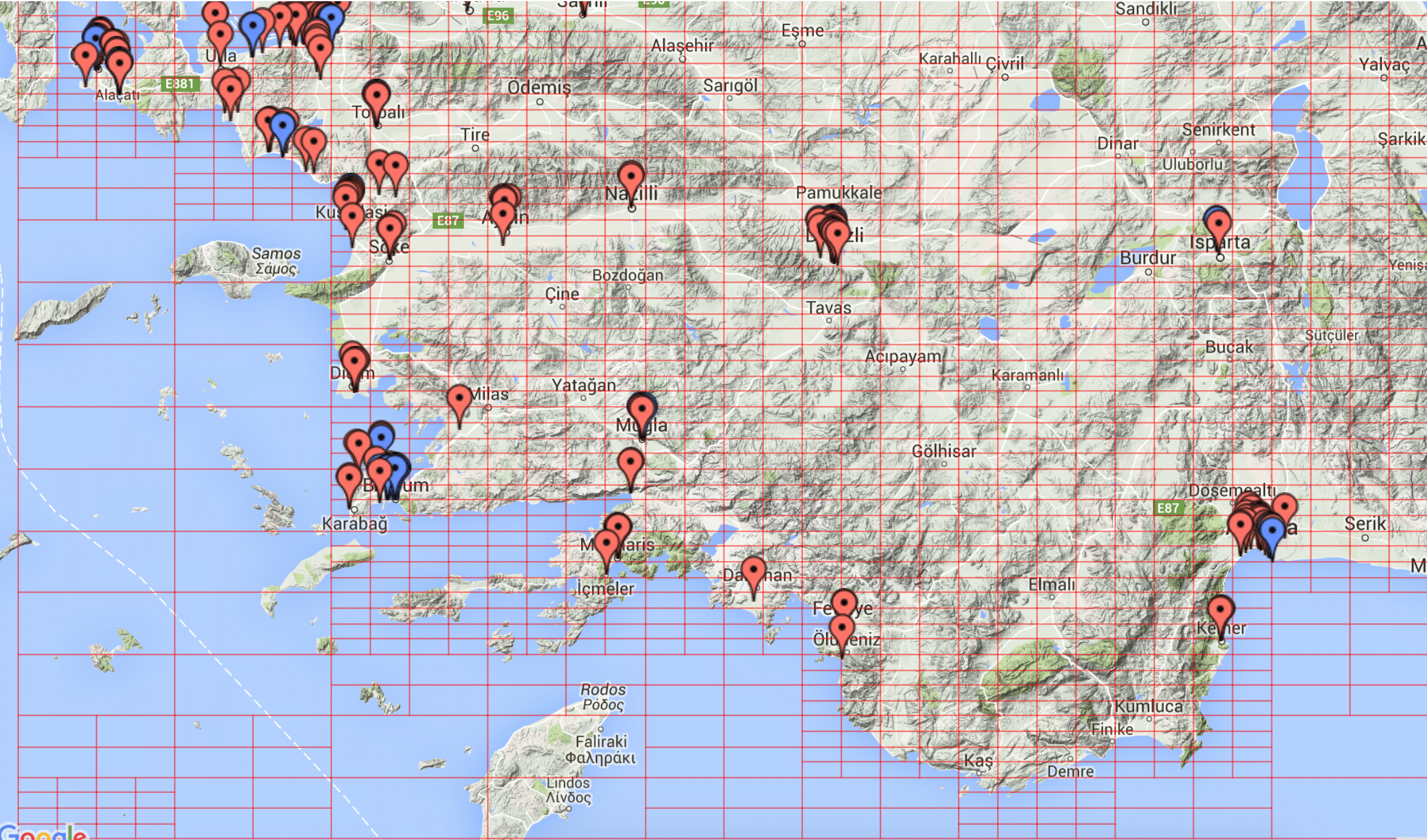}
    \caption{Grid cells.}
    \label{fig:nopoints}
  \end{subfigure}
  \hspace{0.025\linewidth}
  \begin{subfigure}{0.475\linewidth}
    \centering
    \includegraphics[height=0.6\linewidth]{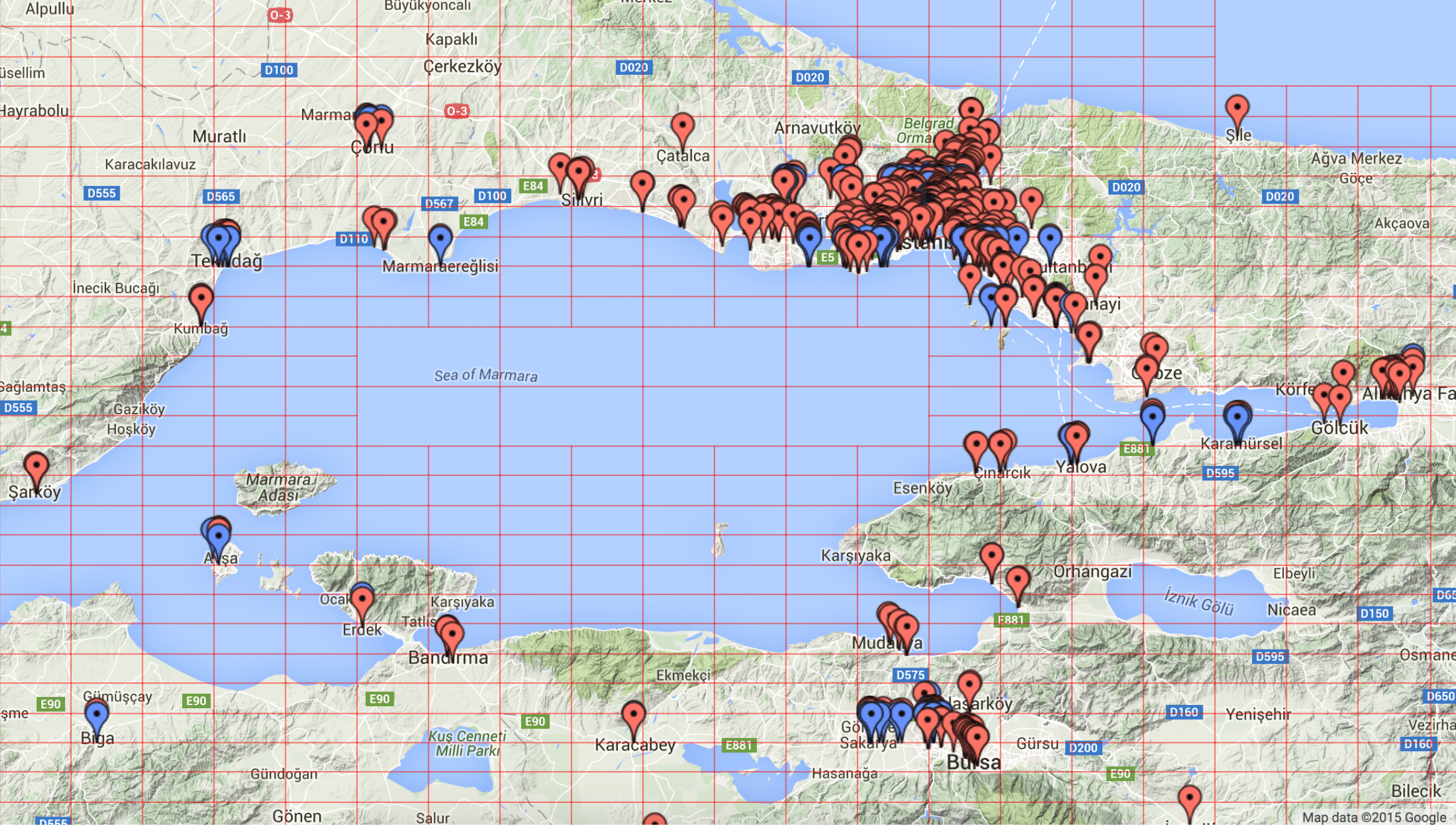}
    \caption{Most popular venues.}
    \label{fig:istanbul}
  \end{subfigure}
  \caption{Grids cells and top 1K venues}
  \label{fig:spatial}
\end{figure}
We also do not split grid cells that do not contain any events. As a result,
not all grid cells are the same size. Figure~\ref{fig:spatial} shows the grid
cells for two selected areas in Turkey and the top 1K venues in those areas in
terms of check-in counts, based on Foursquare check-in data.

\subsubsection{Determining Dominating Grids}

\noindent The determination of the dominating grid for an entity is simply
done by counting the entity's events for different cells and picking the cell with the
highest count. A subtle issue here is about entities whose events end up being
close to the border areas of the grid cells. As a specific example, consider a
user who lives in one cell and works in another. In this case, it is quite
possible that a majority of the user's check-ins happen in one cell and the
majority of the calls in another cell. This will result in missing some of 
the potential matches. To avoid this situation, we make two adjustments:%
\smallskip\\
\indent$\bullet$ If an event is close to the border, then it is counted towards the sums
  for the neighboring cell(s)\footnote{An event can count towards at most 3
  neighbors, in case it is at the corner of the grid.} as well. We use a strip
  around the border of the cell to determine the notion of `close to the
  border'. The width of the strip is taken as the 1/8th of the minimum cell's
  edge width. This means that around $43\%$ of a grid overlaps with one or
  more neighboring grids. This adjustment resembles the loose quad
  trees~\cite{ref:looseTree}.
\smallskip\\
\indent$\bullet$ An entity can potentially have multiple dominating grid cells. We have
  found this to be rare for users in practice.
\smallskip\\
Figure~\ref{fig:istanbul} shows the resulting grids over selected areas in Turkey,
and the most popular venues from our dataset. Red pins are
showing the venues and the blue ones are showing the ones that count towards
neighboring grids.

\subsubsection{Forming Partitioned Datasets}

\noindent Once the dominating grid cells of users are determined, we 
create grid cell specific datasets. For a given grid cell $c$, we take only
the events of the entities who has $c$ as a dominating grid cell. These events
may or may not be in the grid cell $c$. Determination of the dominating grid cells
of entities requires a single scan over the time sorted events from entities. The
forming of the partitioned datasets requires a second scan.

\subsection{Temporal Filtering}
\noindent Temporal filtering aims at creating a small set of candidate user pairs
on which the full linkage algorithm can be executed. To create this set,
temporal filtering looks for user pairs that have co-occurring events, as
expressed by Eq.~\ref{eq:co-occur}. Importantly, temporal filtering also
detects alibi events, based on Eq.~\ref{eq:alibi}, and prevents user
pairs that have such alibi events from taking part in the candidate pair set.

Temporal filtering is based on two main ideas. First, a temporal window is
slided over the events from two different datasets to detect user pairs with
co-occurring events. Since co-occurring events must appear within a given time
duration, the window approach captures all co-occurring events. Second, as the
window slides, alibi events are tracked to prune the candidate user pair
set. However, since the number of alibis is potentially very large, alibis
are only tracked for the user pairs that are currently in the candidate set.
This means that some relevant alibis can be missed if the user pair was added
into the candidate set after an alibi event occurred. To process such alibis
properly, a reverse window scan is performed, during which no new candidate
pairs are added, but only alibis are processed. Algorithm~\ref{algo:slide} gives 
the pseudo-code of temporal filtering.

\begin{algorithm}[!t]
  \begin{small}
  \KwData{$SR_\mathcal{I}$, $SR_\mathcal{E}$: Time sorted datasets of events}
  \KwResult{$CS$: A set of candidate user pairs}
  $CS \leftarrow \varnothing$ \tcp*{Candidates, $CS[u]$ is the list of pair users of $u$}
  $AS \leftarrow \varnothing$ \tcp*{Alibis, $AS[u]$ is the list of alibi users of $u$}
  $UI_x \leftarrow \varnothing, x \in \{ \mathcal{I}, \mathcal{E} \}$ \tcp*{User index over window}
  \tcp{$UI_x[u]$: events from $x$ in window belonging to user $u$}
  $LS_x \leftarrow \varnothing, x \in \{ \mathcal{I}, \mathcal{E} \}$ \tcp*{Spatial index over window}
  \tcp{$LS_x.query(e.r)$: events from $x$ in window intersecting event $e$}
  $W \leftarrow window(SR_\mathcal{I}, SR_\mathcal{E}, \alpha)$ \tcp*{Window over the datasets}
  \tcp{Forward scan phase}
  \While(\tcp*[f]{While more events after window}){$W.hasNext()$}{ 
      \tcp{Get events inserted into and removed from the window}
    $(N^+_\mathcal{I}, N^+_\mathcal{E}, N^-_\mathcal{I}, N^-_\mathcal{E}) \leftarrow W.next()$\\
    \For(\tcp*[f]{In both directions}){$x \in \{\mathcal{I}, \mathcal{E}\}$}{
      \For(\tcp*[f]{For each removed event}){$i \in N^-_x$}{
        $LS_x.remove(i.r, i)$ \tcp*{Remove from spatial index}
        $UI_x[i.u] \leftarrow UI_x[i.u] \setminus i$ \tcp*{Remove from user index}  
      }
    }
    \For(\tcp*[f]{In both directions}){$(x, \bar{x}) \in \{(\mathcal{I}, \mathcal{E}), (\mathcal{E}, \mathcal{I})\}$}{
      \For(\tcp*[f]{For each inserted event}){$i \in N^+_x$}{
        \tcp{Query spatially close elements}
        \For{$e \in LS_{\bar{x}}.query(i.r)$}{
          \If(\tcp*[f]{If users are not alibi}){$e.u \not\in AS[i.u]$}{
            \If(\tcp*[f]{If events co-occur}){$C(i,e)$}{
                \tcp{Add to the candidate set}
              $CS[i.u] \leftarrow CS[i.u] \cup \{e.u\}$\\ 
              $CS[e.u] \leftarrow CS[e.u] \cup \{i.u\}$ 
            }  
          }
        }
        \For(\tcp*[f]{For each candidate user}){$u \in CS[i.u]$}{ 
        \tcp{For each event of the user in the window} 
        \For{$e \in UI_{\bar{x}}[u]$}{
          \If(\tcp*[f]{If $i$ and $e$ is an alibi}){$A(i,e)$}{ 
            \tcp{Add to the alibi set}        
            $AS[i.u] \leftarrow AS[i.u] \cup \{u\}$\\ 
            $AS[u] \leftarrow AS[u] \cup \{i.u\}$\\
            \tcp{Remove from the candidate set}       
            $CS[i.u] \leftarrow CS[i.u] \setminus \{u\}$\\
            $CS[u] \leftarrow CS[u] \setminus \{i.u\}$ 
          }
        }
      }     
      $LS_x.insert(i.r, i)$ \tcp*{Add to spatial index}
      $UI_x[i.u] \leftarrow UI_x[i.u] \cup \{i\}$ \tcp*{Add to user index}
      }
    }
  }
  \tcp{Reverse scan phase}
  $W \leftarrow reverse\_window(SR_\mathcal{I}, SR_\mathcal{E}, \alpha)$ \tcp*{Reverse sliding window}
  \While(\tcp*[f]{While more events after window}){$W.hasNext()$}{
    $(N^+_\mathcal{I}, N^+_\mathcal{E}, N^-_\mathcal{I}, N^-_\mathcal{E}) \leftarrow W.next()$\\
    \For(\tcp*[f]{In both directions}){$x \in \{\mathcal{I}, \mathcal{E}\}$}{
      \For(\tcp*[f]{For each removed event}){$i \in N^-_x$}{
        $UI_x[i.u] \leftarrow UI_x[i.u] \setminus i$ \tcp*{Remove from user index}  
      }
    }
    \For(\tcp*[f]{In both directions}){$(x, \bar{x}) \in \{(\mathcal{I}, \mathcal{E}), (\mathcal{E}, \mathcal{I})\}$}{
      \For(\tcp*[f]{For each inserted event}){$i \in N^+_x$}{
        \For(\tcp*[f]{For each candidate user}){$u \in CS[i.u]$}{ 
          \If(\tcp*[f]{If $i$ and $e$ is an alibi}){$A(i,e)$}{ 
            \tcp{Add to the alibi set}        
            $AS[i.u] \leftarrow AS[i.u] \cup \{u\}$\\ 
            $AS[u] \leftarrow AS[u] \cup \{i.u\}$\\
            \tcp{Remove from the candidate set}       
            $CS[i.u] \leftarrow CS[i.u] \setminus \{u\}$\\
            $CS[u] \leftarrow CS[u] \setminus \{i.u\}$
          }
        }
      }
      $UI_x[i.u] \leftarrow UI_x[i.u] \cup \{i\}$ \tcp*{Add to user index}
    }
  }
  \textbf{return} $CS$ \tcp*{Return the candidate set}
    \caption{Candidate Set Calculation}
    \label{algo:slide}  
  \end{small}   
\end{algorithm}

\begin{figure*}[t]
\centering
\begin{subfigure}{0.54\linewidth}
  \includegraphics[width=\linewidth]{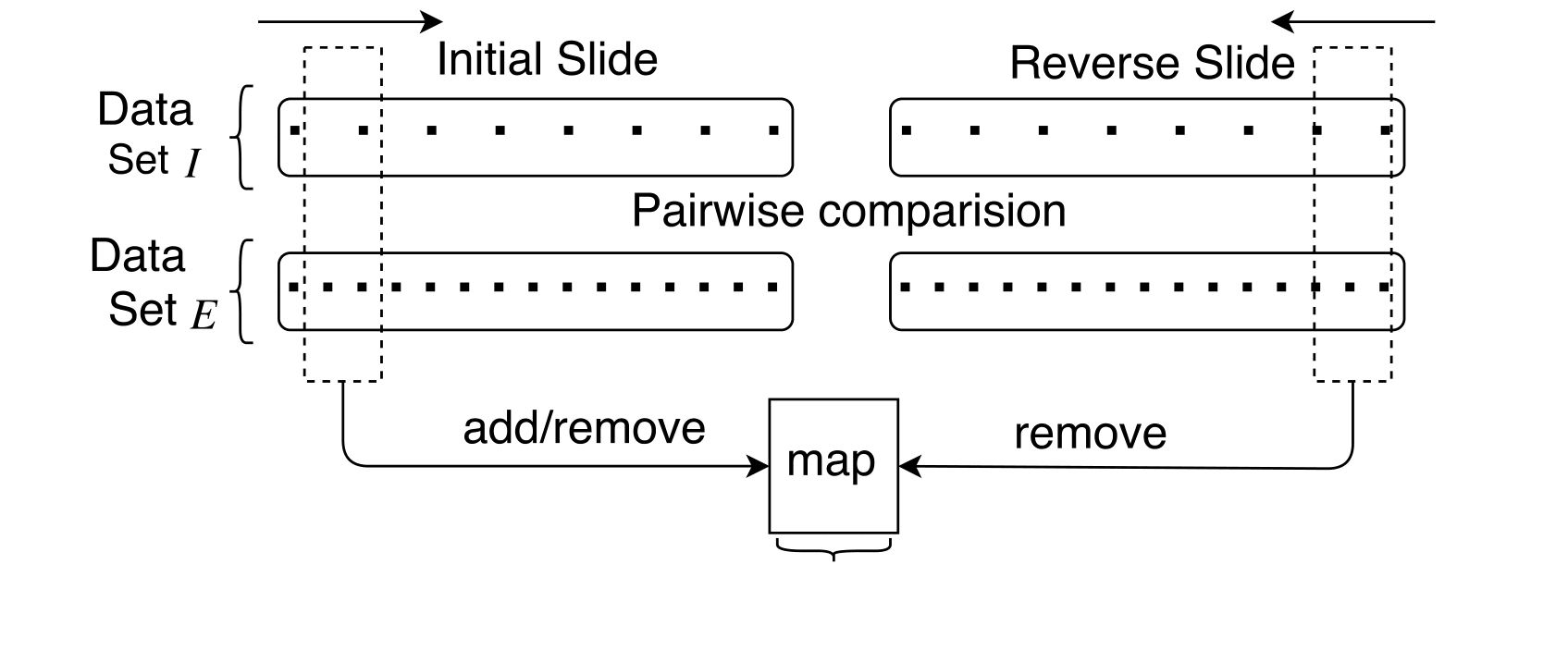}
  \caption{Sliding window algorithm.}\label{fig:sliding_window}
\end{subfigure}
\hspace{0.01\linewidth}
\begin{subfigure}{0.42\linewidth}
  \includegraphics[width=\linewidth]{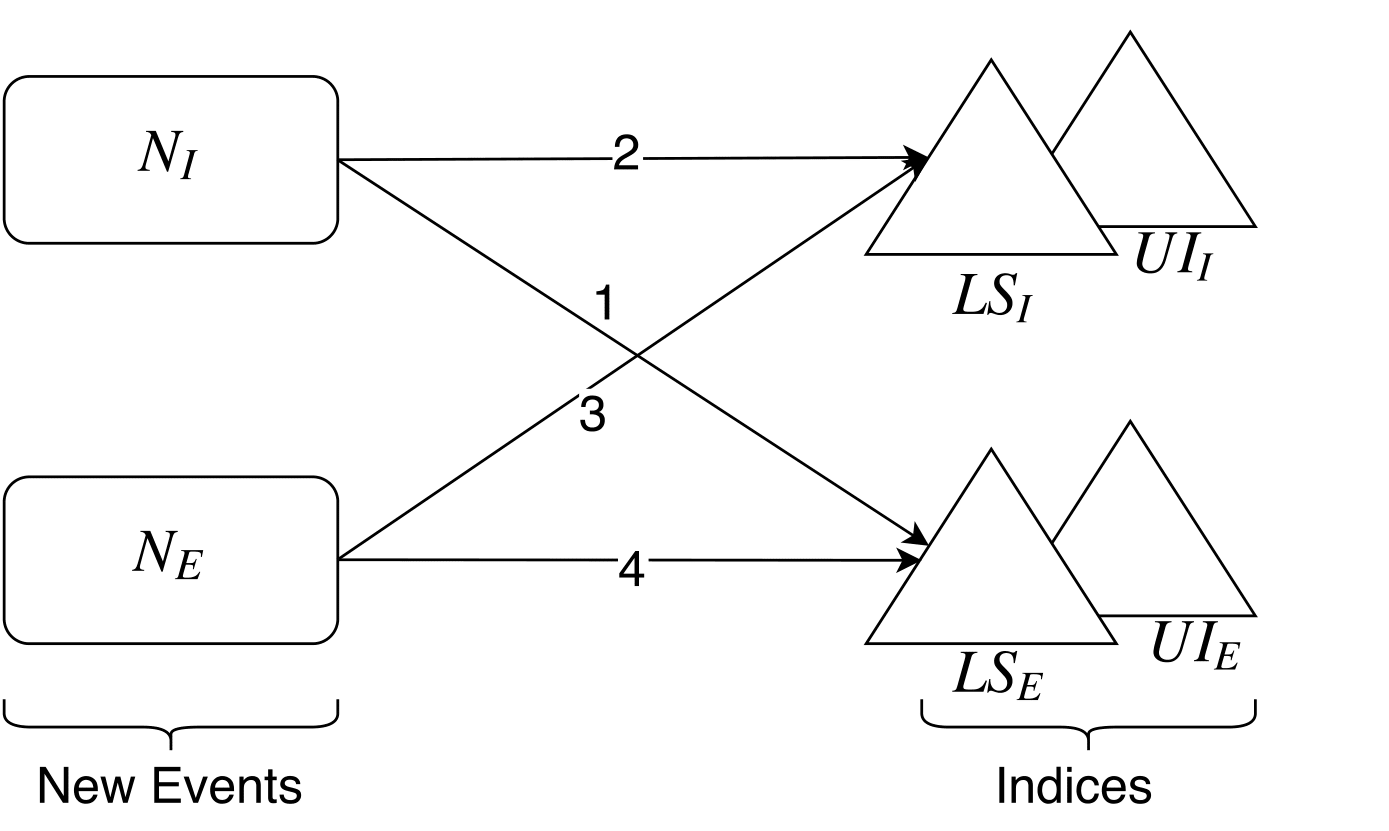}
  \caption{Calculation of the candidate set.}\label{fig:algo}
\end{subfigure}
\caption{Temporal Filtering}\label{fig:temporal}
\end{figure*}

\subsubsection{Data Structures}
\noindent 
A window of size $\alpha$ (see Eq.~\ref{ref:co-time}) is slided jointly over both
time sorted datasets.
Figure~\ref{fig:sliding_window} depicts this visually. Each time the window
slides, some events from both datasets may enter and exit the window. We utilize
two types of data structures to index the events that are currently in the
window.
The first type of index we keep is called the \emph{user index}, denoted by
$UI_x$, where $x\in\{\mathcal{I}, \mathcal{E}\}$. In other words, we keep
separate user indexes for the two datasets. $UI_x$ is a hash map indexed by
the user. $UI_x[u]$ keeps all the events (from dataset $x$) of user $u$ in
the window. As we will see, this index is useful for quickly checking alibis.

The second type of index we keep is called the \emph{spatial index}, denoted
by $LS_x$, where $x\in\{\mathcal{I}, \mathcal{E}\}$. Again, we keep separate
indexes for the two datasets. $LS_x$ could be any spatial data structure like R-trees.
$LS_x.query(r)$ gives all events whose region intersect with
region $r$. As we will see, this index is useful for quickly locating 
co-occurring events.

In addition to these indexes, we maintain a global candidate set $CS$ and a
global alibi set $AS$. For a user $u$ (from either dataset, assuming user ids
are unique), $CS[u]$ keeps the current set of candidate pair users for $u$;
and $AS[u]$ keeps the current known alibis users for $u$. It is important to
note that $AS$ is not designed to be exhaustive. For a user $u$, $AS[u]$ only
keeps alibi users that have co-occurring events with $u$ in the dataset.

\subsubsection{Processing Window Events}
\noindent
The algorithm operates by reacting to events being inserted and removed from
the window as the window slides over the dataset. As a result, an 
outermost while loop that advances the window until the entire dataset is
processed. At each iteration, we get a list of events inserted
($N^+_\mathcal{I}$ and $N^+_\mathcal{E}$) and removed ($N^-_\mathcal{I}$ and
$N^-_\mathcal{E}$) from the window. We first process the removed events, which
consists of removing them from the spatial and user indexes. We then process
the inserted events. We first process $N^+_\mathcal{I}$ against
$UI_{\mathcal{E}}$ and $LS_{\mathcal{E}}$, then insert the events in
$N^+_\mathcal{I}$ into $UI_{\mathcal{I}}$ and $LS_{\mathcal{I}}$, then process
$N^+_{\mathcal{E}}$ against $UI_{\mathcal{I}}$ and $LS_{\mathcal{I}}$, and
finally  insert the events in $N^+_{\mathcal{E}}$ into $UI_{\mathcal{E}}$ and
$LS_{\mathcal{E}}$. This ensures that all the events are compared, and no
repeated comparisons are made. Figure~\ref{fig:algo} depicts the order of
events visually.

To compare a new event $i$ from dataset $x$ against the events from dataset
$\bar{x}$ that are already indexed in the window (where $\{x, \bar{x}\} =
\{\mathcal{E}, \mathcal{I}\}$), we use the indexes $UI_{\bar{x}}$ and $LS_{\bar{x}}$. 
First, we find events that co-occur with $i$ by considering events $e$ in $LS_{\bar{x}}.query(i.r)$. 
These are events whose regions intersect with that of
$i$. If the user of such an event $e$ is not already a known alibi of the user
of $i$ (not in $AS[i.u]$) and if the co-occurrence condition $C(i, e)$ is
satisfied, then the user $e.u$ and user $i.u$ are added as candidate pairs of
each other. Second, and after all the co-occurrences are processed, we
consider all candidate users of the event $i$'s user, that is $CS[i.u]$, for
alibi processing. For each user $u$ in this set, we check if any of its events
result in an alibi. To do this, we iterate over user $u$'s events with the
help of the index $UI_{\bar{x}}$. In particular, for each event $e$ in
$UI_{\bar{x}}[u]$, we check if $i$ and $e$ are alibis, using the condition
$A(i, e)$. If they are alibis, then we remove $u$ and $i$'s user ($i.u$) from
each other's candidate sets, and add them to their alibi sets.

This completes the description of the forward scan of the window. An important
point to note is that, during the forward window scan, we only check alibis
for user pairs that are in the set of candidate pairs. It is possible that
there exists an alibi event pair for users $x$ and $y$, that appears before
the first co-occurring event pair for these users. In such a case, during the
processing of the alibi events we won't have this pair of users in our
candidate set and thus their alibi will be missed. To fix this problem, we
perform a reverse scan. During the reverse scan, we only process alibis, as no
new candidate pairs can appear. Furthermore, we need to process alibi events
for a user pair only if the events happened before the time this pair was
added into the candidate set. For brevity, we do not show this detail in
Algorithm~\ref{algo:slide}. At the end of the reverse scan, the set $CS$
contains our final candidate user pairs, which are sent to the linkage step.
Temporal filtering is highly effective in reducing the number of pairs for
which complete linkage procedure is executed. The experimental results show the
effectiveness of this filtering.

When there is inaccurate information in the datasets,
disregarding candidate pairs due to only a single alibi event might lead to
false negatives. However, the algorithm is easily modifiable to use a
threshold for alibi values. In this modified version, we update the structure
of the alibi set $AS$ to keep the number of alibi events of a pair as well.
Now $AS[u]$ keeps the current known alibi users of user $u$ with alibi event
counts for each. Just like in the original algorithm, when two events $i$ and
$e$ are compared we first check if the number of alibi events of users $i.u$
and $e.u$ exceeds the threshold. To avoid double counting, we reset
the counters before the reverse scan. Since all alibi events of current
candidate pairs will be counted in reverse scan, candidates whose count of
alibi events exceed threshold will not be included in the resulting candidate
set $CS$.

So far we have operated on time sorted event data and our algorithms used only
sequential I/O. However, during the linkage step, when we finally decide
whether a candidate user pair can be linked, we will need the time sorted
events of the users at hand. For that purpose, during the forward scan, we
also create a disk-based index sorted by the user id and event time. This
index enables us to quickly iterate over the events of a given user in
timestamp order, which is an operation used by the linkage step. For this
purpose, we use LevelDB~\cite{ref:leveldb} as an index, which is a
log-structured merge-tree supporting fast insertions.

While writing the event to the disk-based index, we also include information
about the number of unique users the event has matched throughout its stay in
the forward scan window. This information is used as part of the weight
calculation (recall Eq.~\ref{eq:weight}) in the linkage step.

\subsubsection{Handling Time Period in Events}\label{sec:timePeriod}
The temporal filtering step scans time-ordered events by sliding a window of
size $\alpha$ over them. This operation assumes that the time information is a
point in time. Yet, there could be scenarios where the time information is a
period (e.g., a start time and a duration). However, frequently, these records
contain only start location of the event. For example, although Call Detail
Records (CDR) have the start time of the call and the duration, they usually
contain only the originating cell tower information. Considering mobility of
the users, assuming a fixed location during this period would lead to location
ambiguity.

If we have events with time periods and accurate location information is
present during this period, we can adapt our approach to handle this. In
particular, we need to avoid false negative candidate pairs when the event
contains a time period. Since events are processed via windowing, making sure
that the event with the time period information stays in the window as long as
its time period is valid would guarantee that all co-occurrences will be
processed. This requires creating multiple events out of the original event,
with time information converted into a point in time and the correct location
information attached to it. The number of such events is bounded by the time
duration divided by the window size, $\alpha$.

\subsection{Linkage}
\label{sec:linkage}
\noindent The last step of the \emph{ST-Link} algorithm is the linkage of the
entities that are determined as possible pairs as a result of spatial and
temporal filtering. This linkage is a realization of the $k$-$l$ diversity
based linkage model introduced in Section~\ref{sec:formalization}. Given two
entities from two datasets, the linkage step uses the events of them to
determine whether they can be linked according to Eq.~\ref{eq:k-l-match}.
Thanks to efficient filtering steps applied on the data beforehand, the number
of entity pairs for which this linkage computation is to be performed is
significantly reduced.

For each entity pair, their events are retrieved from the disk-based index
created as part of the forward scan during the temporal filtering. These
events are compared for detecting co-occurring events. Co-occurring events are
used to compute the $k$ value, via simple accumulation of the co-occurrence
weights. They are also used to accumulate weights for the places where
co-occurring events occur. This helps us compute the $l$ value, that is
diversity. After all events of a pair of entities are compared, we check if they
satisfy the $k$-$l$ diversity requirement. Note that, it is not possible to
see an alibi pair event at this step, as they are eliminated by the temporal
filtering step.

There are a number of challenges in applying the $k$-$l$-diversity based
linkage. The first is to minimize the number of queries made to the disk-based
index to decrease the I/O cost. Events from the same entity are stored in
a timestamped order within the index, which makes this access more efficient.
Also, if one of the datasets is more sparse than the other, then the linkage
can be performed by iterating over the entities of the dense datasets first,
making sure their events are loaded only once. This is akin to the classical
join ordering heuristic in databases.

Another challenge is the definition of the place ids to keep track of
diversity. A place id might be a venue id for a Foursquare dataset, store id
for credit card payment records, cell tower id for Call Detail Records, or a
geographic location represented as latitude and longitude. An important
difference is the area of coverage for these places. Consider two datasets of
Foursquare check-ins and Call Detail Records, and places based on venues. If a
user visits several nearby coffee shops and makes check-ins and calls, these
will be considered as diverse even though they are not geographically diverse.
The use of cell tower coverage areas is a more practical choice for
determining places.

The last challenge is about matching events. Recall from
Figure~\ref{fig:match} that events of two entities can be matched in multiple
different ways, resulting in different weights for the co-occurrences. Ideally,
we want to maximize the overall total weight of the matching, however this
would be quite costly to compute, as the problem is a variation of the
\emph{bipartite graph assignment problem}. As a result, we use a greedy
heuristic. We process events in a timestamped order and match them to the
co-occurring event from the other entity that provides the highest weight. Once
a match is made, event pairs are removed from the dataset so that they are not
re-used. 

Different $k$-$l$ value pairs may perform significantly different in terms of
precision and recall, depending on the frequencies of the events in the
datasets. An ad-hoc approach is to decide the $k$ and $l$ values based on
observation of results from multiple experimental runs. A more robust
technique we used is to detect the best \emph{trade-off} point (a.k.a
\emph{elbow point}) on a curve. Given the \emph{co-occurrence} and
\emph{diversity} distributions, we independently detect the elbow point of
each, and set the $k$ and $l$ values accordingly. Although there is no
unambiguous solution for detecting an elbow point, the maximum absolute second
derivative is an approximation. Let $A$ be an array of co-occurrence (or
diversity) values with size $n$. Second derivative, $SD$, of point at index
$i$ can be approximated with a central difference as follows:
\begin{equation}
  SD[i] = A[i+1] + A[i-1] - 2* A[i]
\end{equation}

The value at index $A[i]$, such that $i$ has the maximum
absolute $SD[i]$ value, is selected as the \emph{elbow point} and $k$ (or $l$) 
value is set accordingly.

\section{Experimental Evaluation}\label{sec:experiments}
\begin{figure*}[t]
\centering
\begin{subfigure}{.30\linewidth}
  \includegraphics[width=\linewidth]{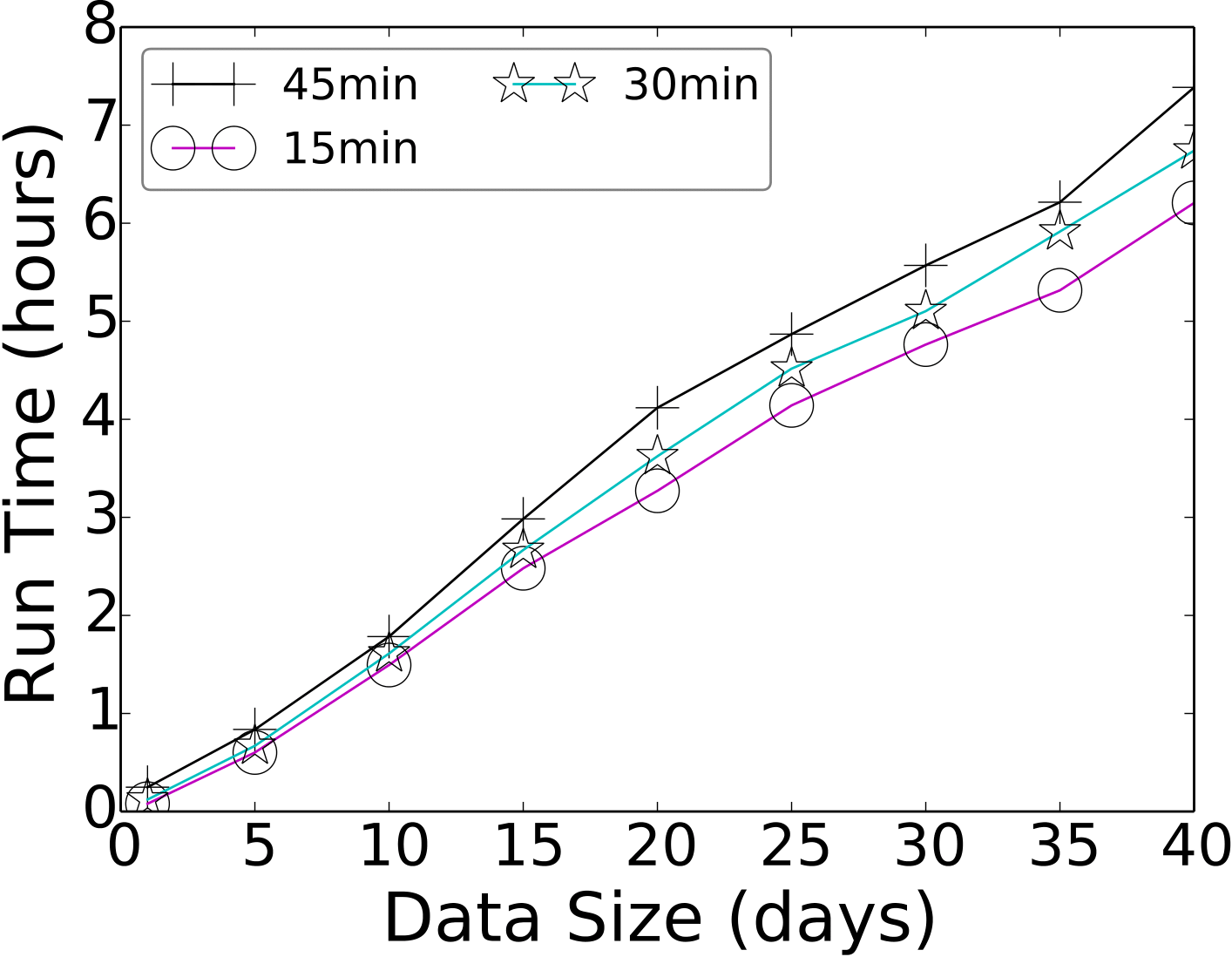}
  \caption{Running time vs. dataset size.\hfill\eject$\,$}
  \label{fig:runtime}
\end{subfigure}
\hspace{0.01\linewidth}
\begin{subfigure}{0.30\linewidth}
  \includegraphics[width=\linewidth]{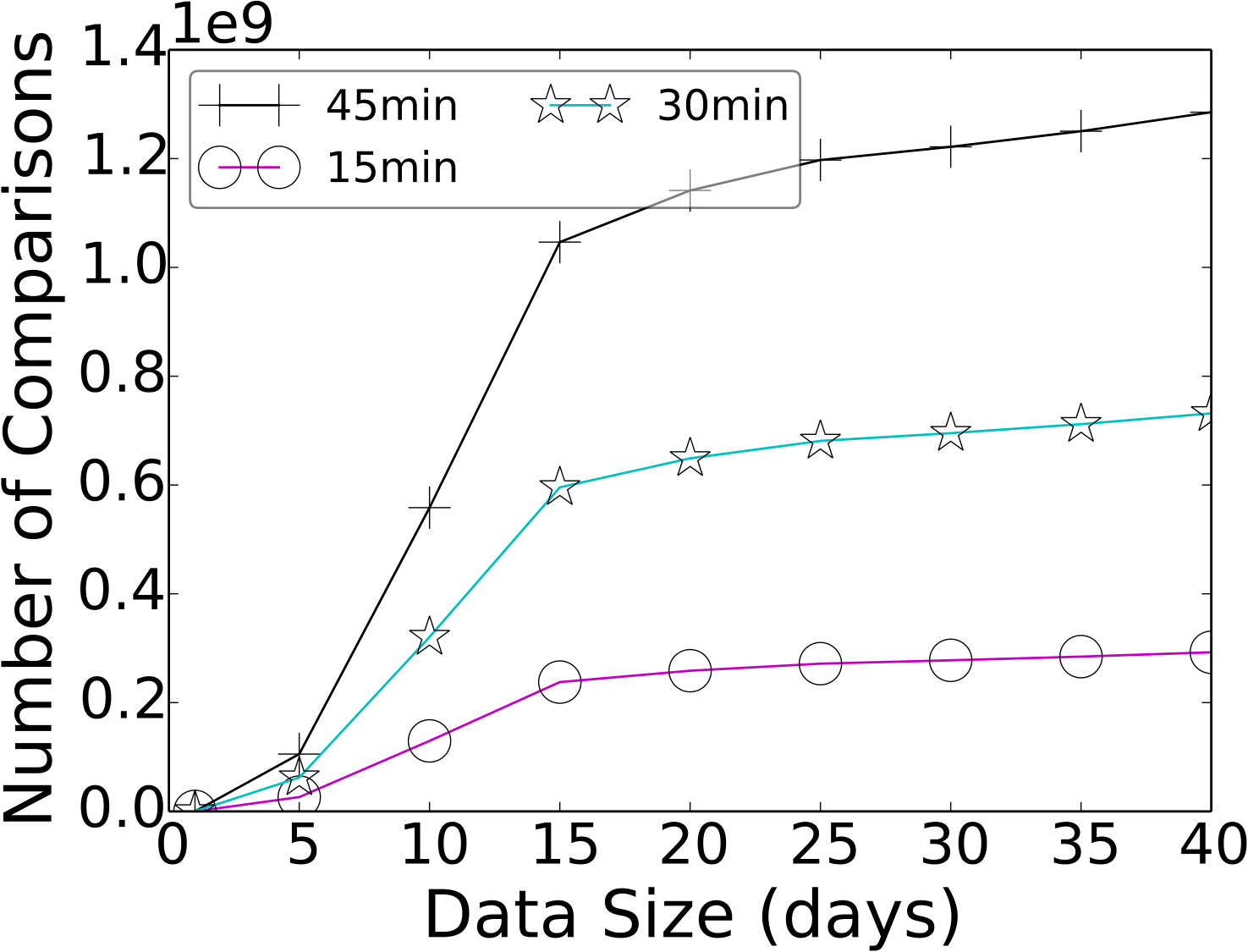}
  \caption{Number of comparisons vs. dataset size.}
  \label{fig:comparisons}
\end{subfigure}
\hspace{0.01\linewidth}
\begin{subfigure}{.30\linewidth}
  \includegraphics[width=\linewidth]{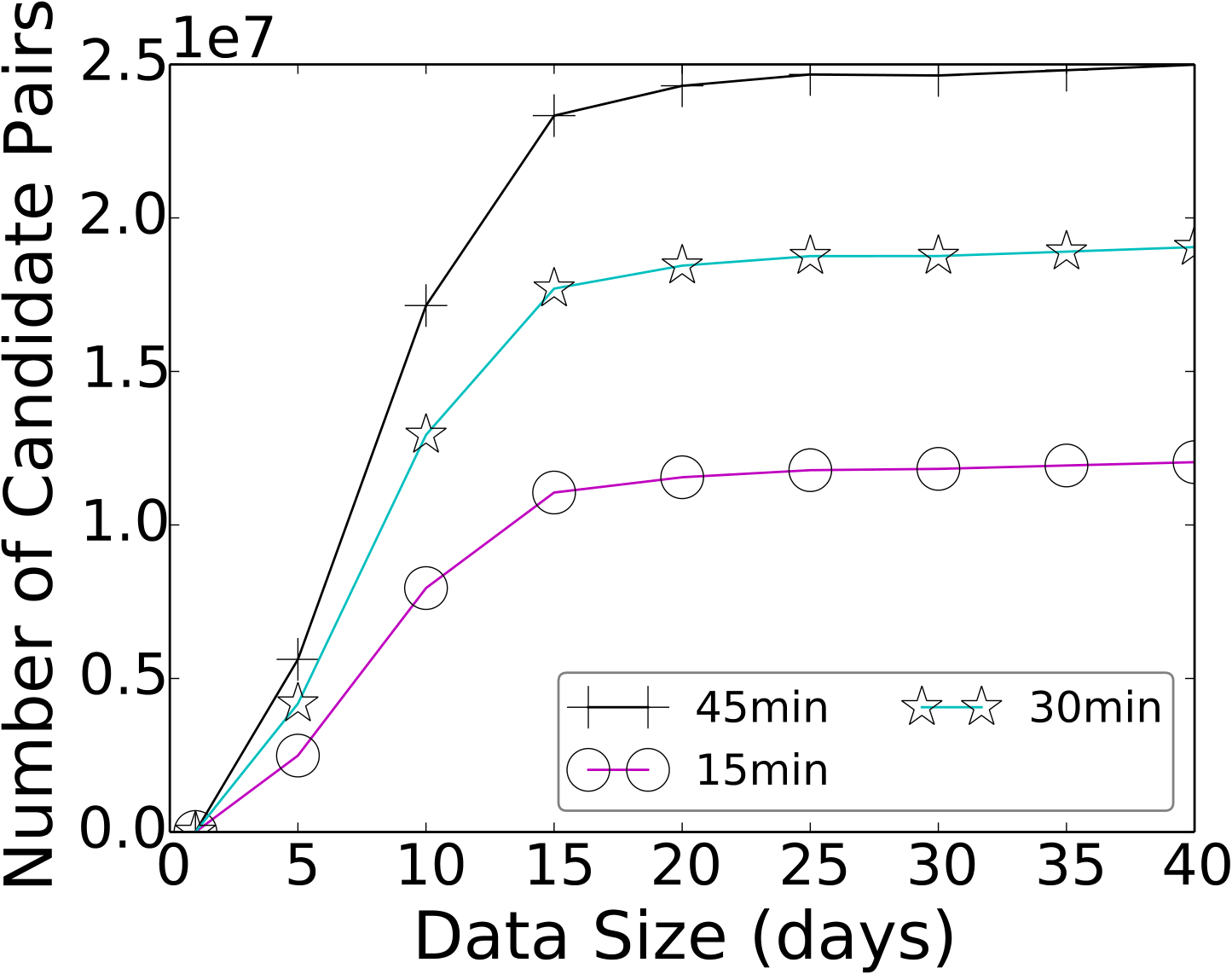}
  \caption{Number of candidate user pairs vs. dataset size.}
  \label{fig:pairs}
\end{subfigure}
\caption{Performance Results}
\label{fig:performance}
\end{figure*}

\noindent In this section, we present an evaluation of the
proposed $k$-$l$ diversity based linkage method and the
\emph{ST-Link} algorithm. We implemented the \emph{ST-Link} algorithm
using Java 1.7. All experiments were executed on a Linux server with $2$
Intel Xeon E5520 $2.27$GHz CPUs and $64$GB of RAM. 

We present two sets of experiments. In the first set of
experiments, we measure the performance and the scalability of the
\emph{ST-Link} algorithm. By increasing the size of the input data, we 
test the change in the running time, number of event comparisons, and the
number of candidate user pairs, for different window sizes.  In the second set of experiments, we analyze the quality of the $k$-$l$ diversity based linkage. To measure quality, we use two metrics. The first is the
precision, which measures the fraction of correctly linked pairs in the list
of user pairs produced by \emph{ST-Link}. The second is the number of true
positives, which is the number of user pairs correctly linked by the
\emph{ST-Link} algorithm. 

\subsection{Datasets Used}\noindent
For the performance, scalability, and accuracy evaluations we used three datasets. The first is a Foursquare dataset
of check-ins. The second is anonymized call detail records in a telecommunication
provider. For privacy concerns, we did not perform any linkage across these two datasets. 
As a result, we were not able to compute accuracy results when using these two datasets.
However, they are used for the evaluation of running time performance. 
To evaluate accuracy, linkage is performed
between a third dataset belonging to a hypothetical LES and the call dataset. This dataset was synthetically
derived to protect privacy, from the call dataset by ($i$) picking a predefined fraction $f$ of the
callers at random as active users of the second LES, ($ii$) generating usage
records for the selected users by assuming that they generate such a record
with probability $p$, within a $15$ minute time window of a call, inside a
location within the same cell tower of the call. We change the parameters $p$
and $f$ to experiment with different scenarios. Lower values for $p$ result in
a sparser usage record dataset for the second LES. We call the parameter $p$,
the \emph{check-in probability}. As not all users have the same check-in
probability in practice, we pick the value of the check-in probability for a given user
from a Gaussian distribution with mean $p$. We call the parameter $f$, the
\emph{usage ratio}.
\begin{table}[h!t] 
\centering
\begin{tabular}{|r||c|c|}\hline 
\emph{Datasets $\Rightarrow$}  
                            & Foursquare   & Call          \\\hline\hline  
\# of activities          & 1,903,674    & 1,890,107,057 \\\hline 
\# of venues/cell towers    & 300,685      & 109,780       \\\hline 
\# of users         & 284,856      & 3,357,069     \\\hline
\end{tabular}
\caption{Dataset statistics}\label{tbl:data}
\end{table}  

The Foursquare dataset consists of check-ins that were shared publicly on
Twitter, collected via the Twitter streaming
API\footnote{www.dev.twitter.com} and the Foursquare
API\footnote{www.developer.foursquare.com}. This dataset spans $40$ days and
only contains check-ins from Country X. Each row contains the acting user's
Foursquare id, venue id, geographical location (lat/lon) of the
venue, and the time of the check-in. The call dataset spans 
the same $40$ days in Country X. Each row contains an anonymized id, time of the call, and geographical location
(lat/lon) of the handling cell tower. The anonymized id is the same across all usage
of the same user. Table~\ref{tbl:data} shows the statistics about both the Foursquare and the
call datasets. For the runtime performance and filtering effectiveness experiments, we used
the two real datasets. However, since it is not possible to
verify the accuracy of the results using these two datasets, we used the synthetic dataset which is derived from the
anonymous call data for the evaluation of \emph{ST-Link}'s accuracy.

\subsection{Running Time Performance}
\begin{figure*}[t]
\centering
\begin{minipage}{.30\linewidth}
  \includegraphics[width=\linewidth]{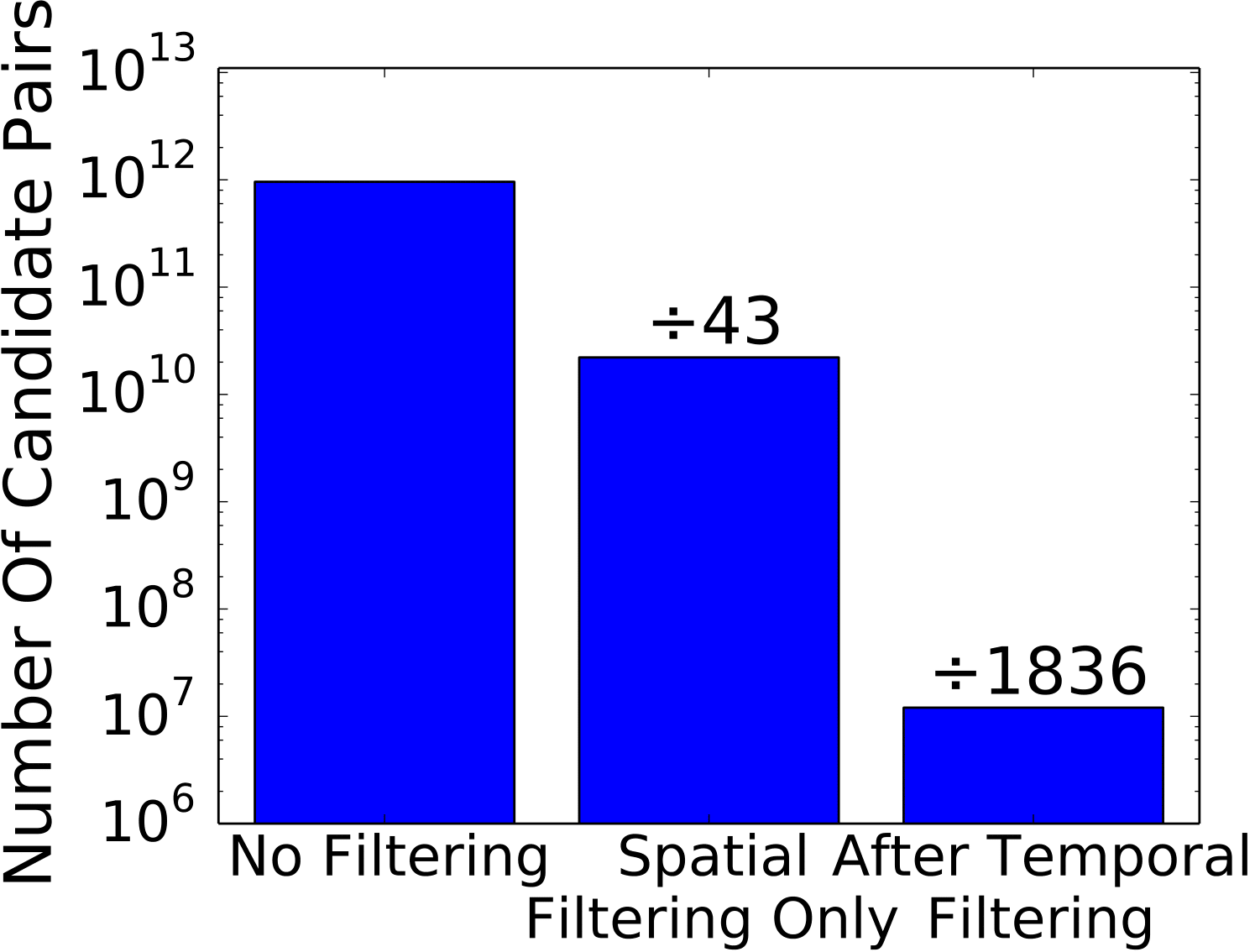}
    \caption{Reduction in the number of possible pairs.}
    \label{fig:efficiency}
\end{minipage}
\hspace{0.01\linewidth}
\begin{minipage}{0.30\linewidth}
  \includegraphics[width=\linewidth]{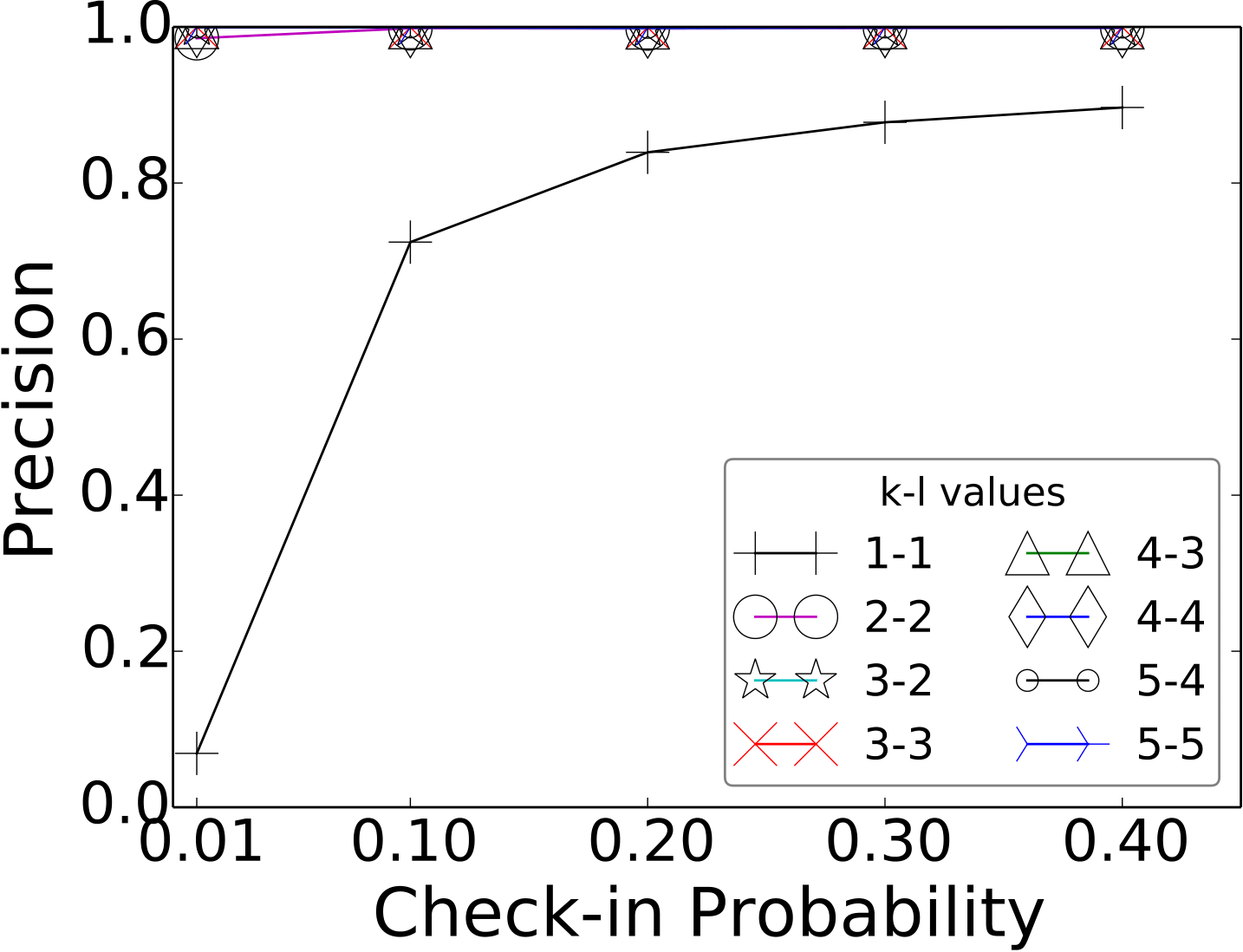}
  \caption{Precision as a function of check-in probability.}
  \label{fig:meanPrecision}
\end{minipage}
\hspace{0.01\linewidth}
\begin{minipage}{0.30\linewidth}
  \includegraphics[width=\linewidth]{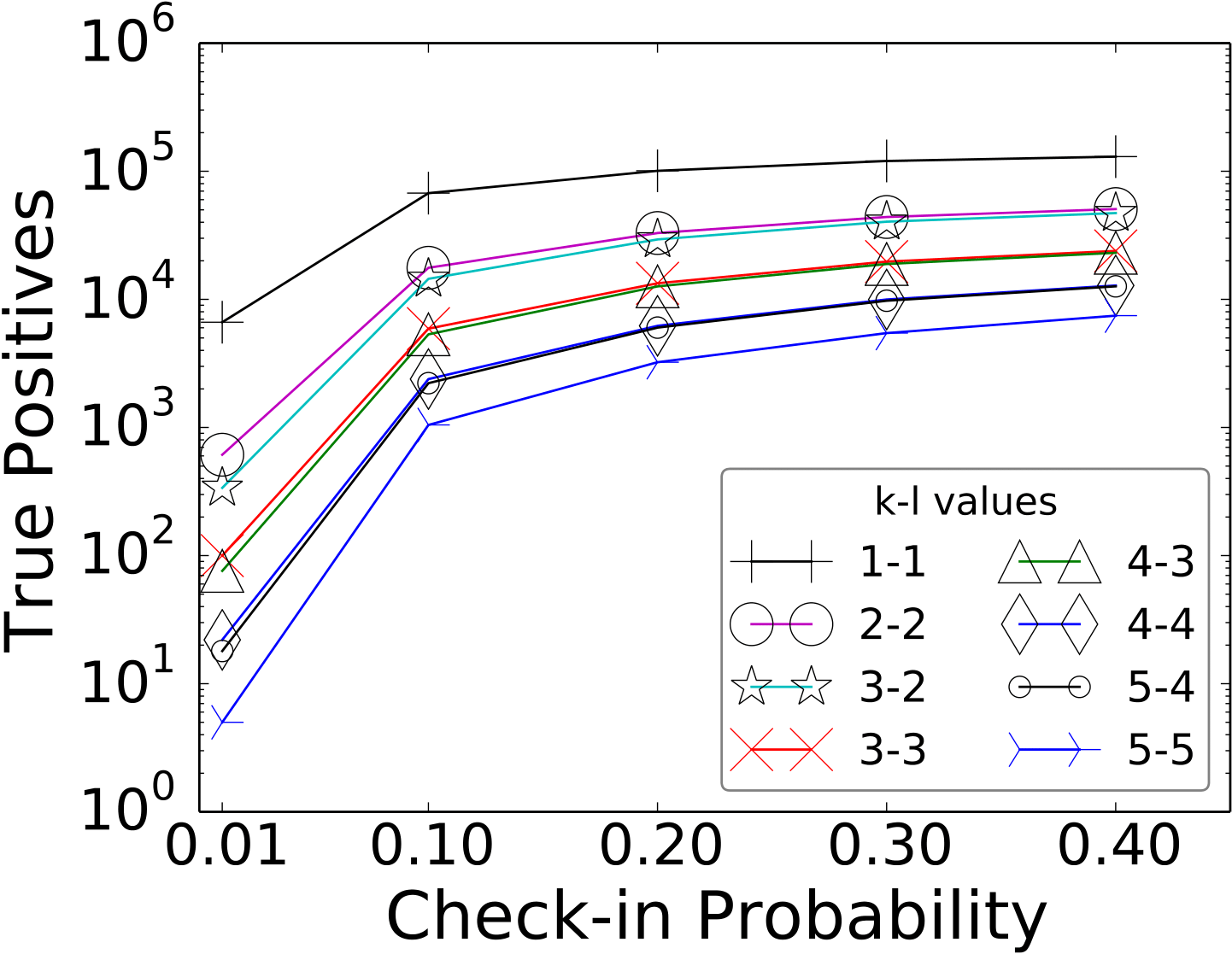}
  \caption{Number of true positives as a function of check-in probability.}
  \label{fig:meanTP}
\end{minipage}
\end{figure*}

\begin{figure*}[t]
\centering
\begin{minipage}{0.30\linewidth}
    \includegraphics[width=\linewidth]{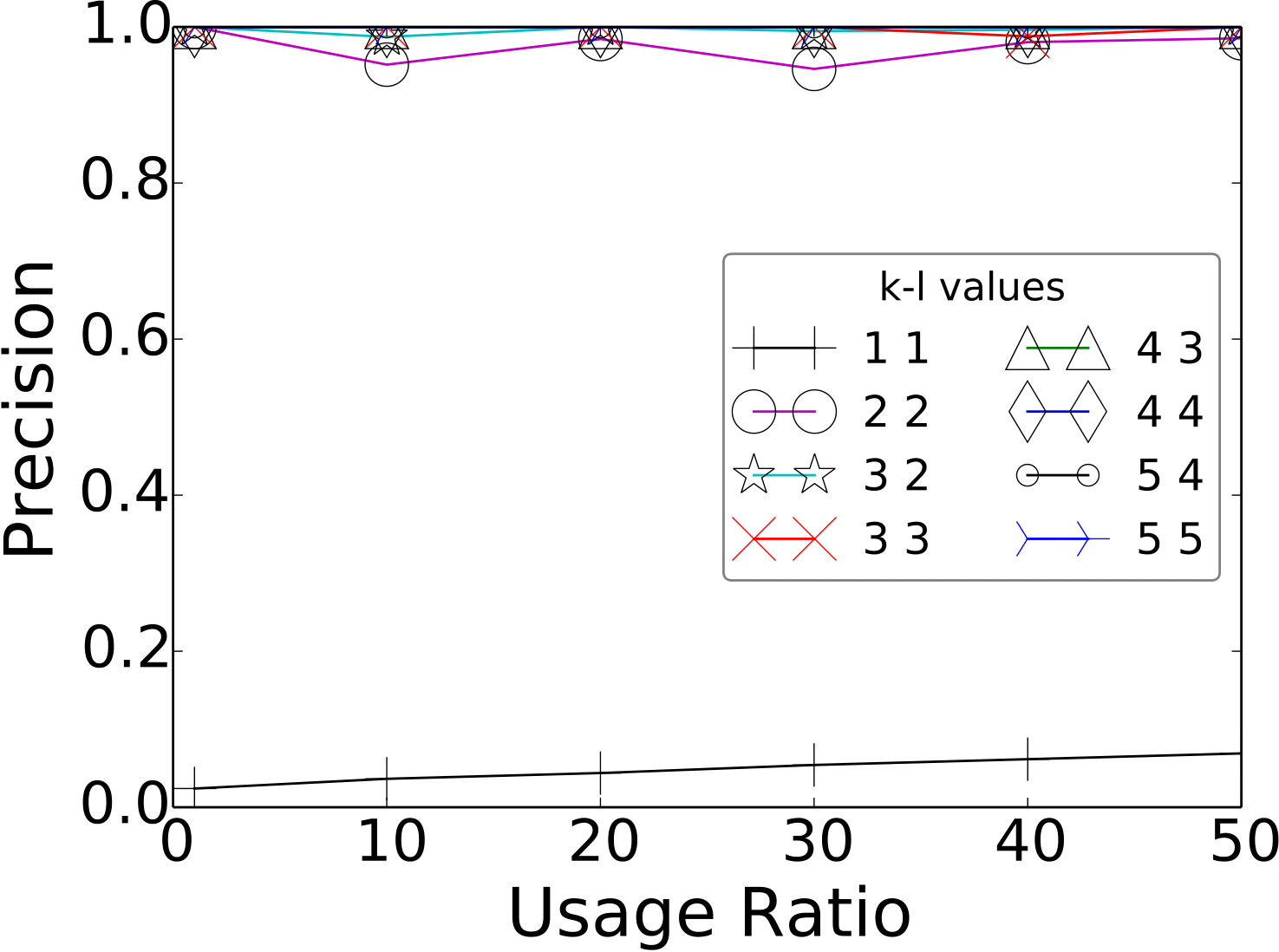}
    \caption{Precision as a function of usage ratio.}
    \label{fig:activePrecision}
\end{minipage}
\hspace{0.01\linewidth}
\begin{minipage}{0.30\linewidth}
  \includegraphics[width=\linewidth]{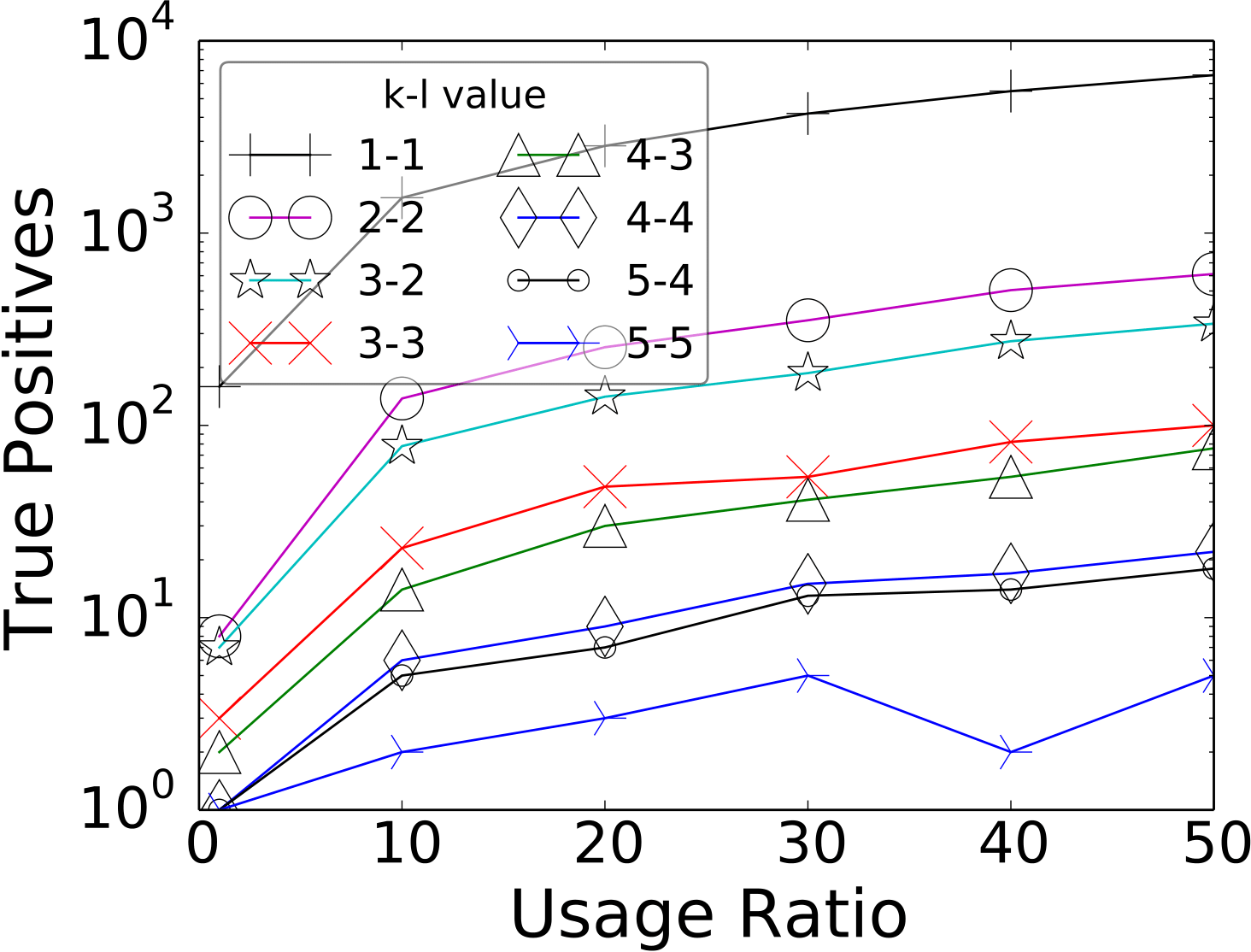}
  \caption{Number of true positives as a function of usage ratio.}
  \label{fig:activeTP}
\end{minipage}
\hspace{0.01\linewidth}
\begin{minipage}{0.30\linewidth}
\centering
  \begin{tabular}{|r||c|c|}\hline 
          & Precision   & Recall \\ \hline 
    2-2   & 0.19     & 0.31   \\ \hline 
    3-2   & 0.36      & 0.32  \\ \hline
\rowcolor{Gray}   3-3   & 0.89    & 0.61 \\ \hline
    4-3   & 0.93    & 0.47 \\ \hline
    4-4   & 0.99    & 0.58 \\ \hline
    5-4   & 0.99    & 0.47 \\ \hline
    5-5   & 0.99    & 0.50 \\ \hline
  \end{tabular}
  \caption{Precision and recall using unweighted linkage}
  \label{tbl:weightTest}
\end{minipage}
\end{figure*}
\begin{figure*}[t]
\centering
\begin{minipage}{0.30\linewidth}
    \includegraphics[width=\linewidth]{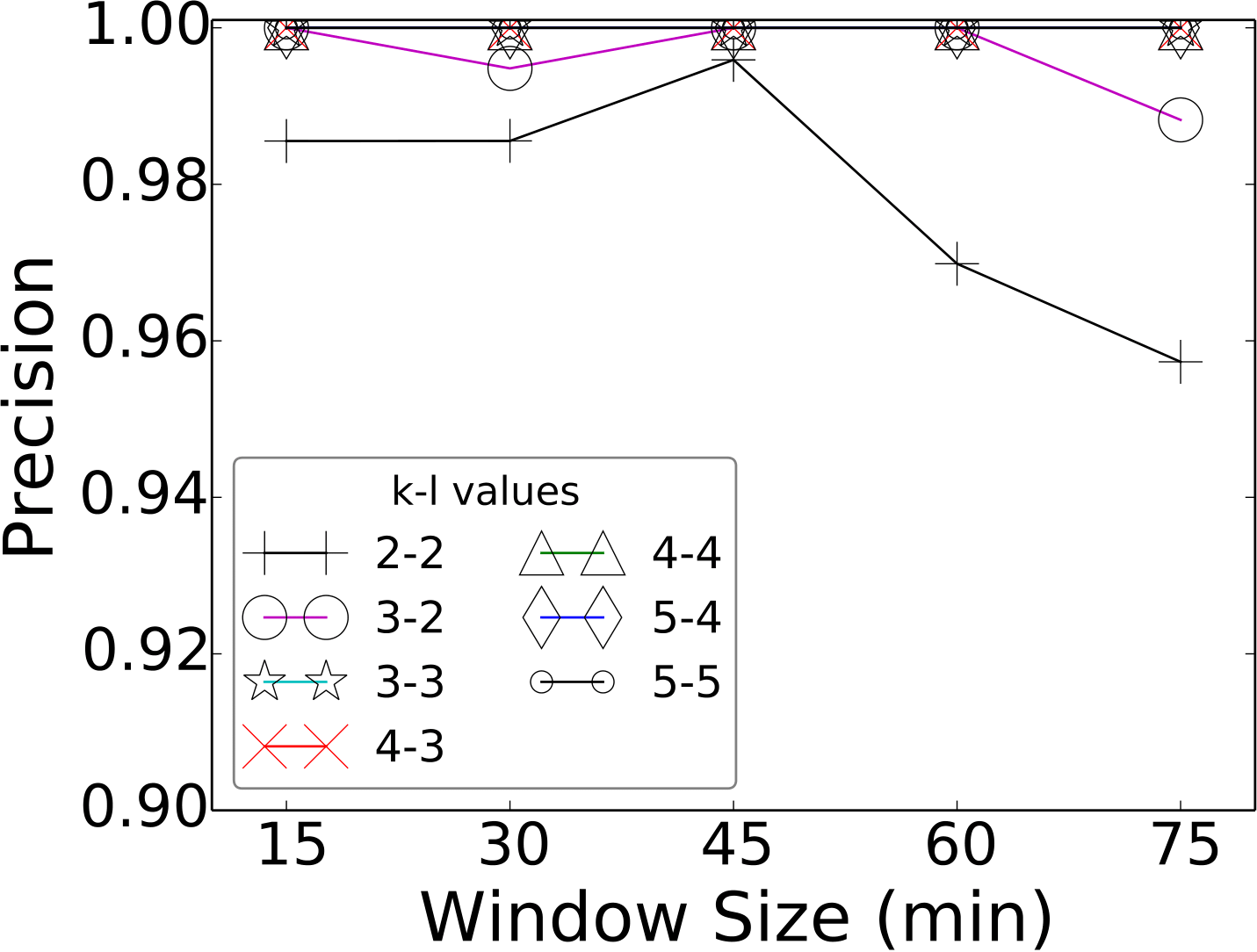}
    \caption{Precision as a function of window size.}
    \label{fig:wsPrecision}
\end{minipage}
\hspace{0.01\linewidth}
\begin{minipage}{0.30\linewidth}
  \includegraphics[width=\linewidth]{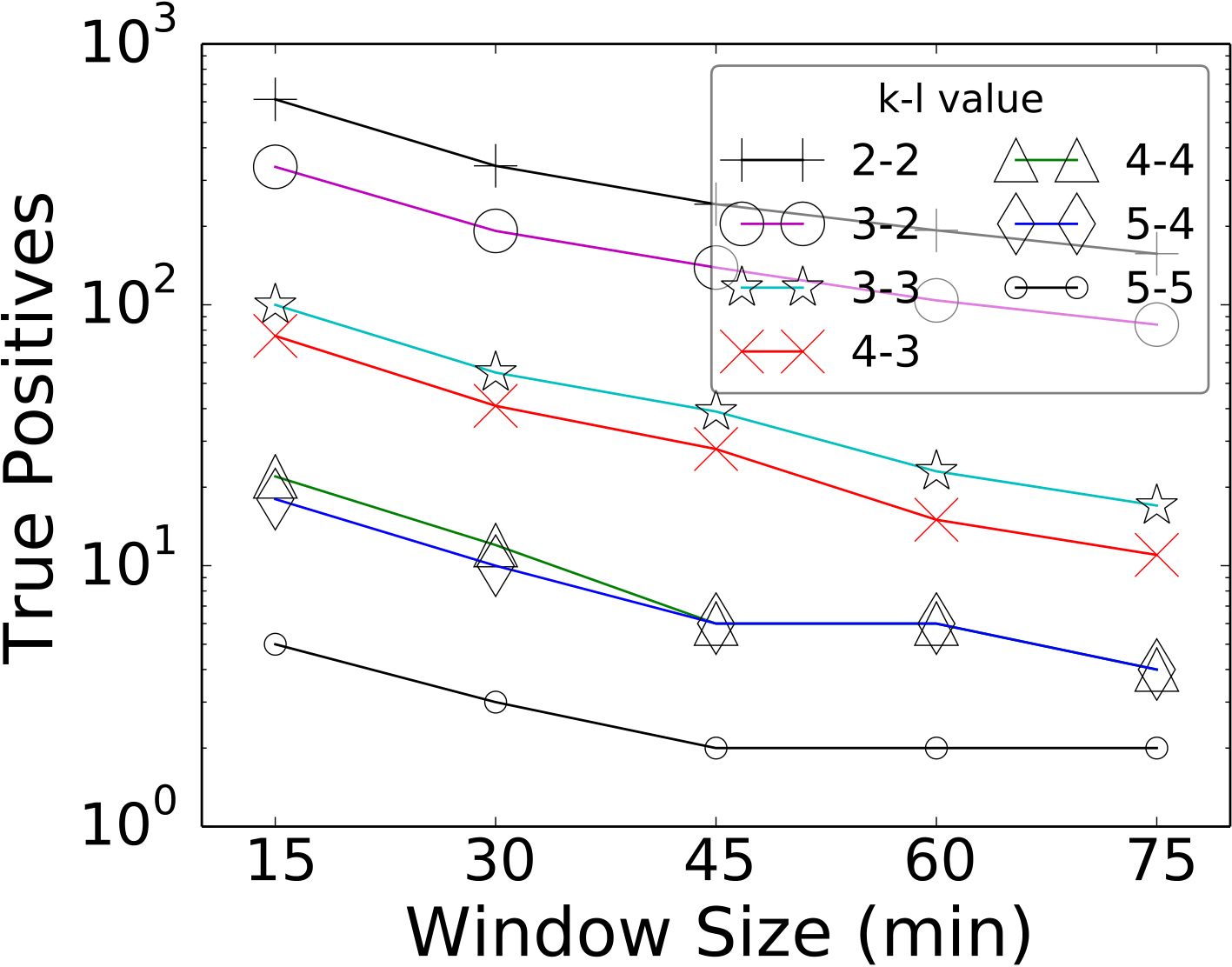}
  \caption{Number of true positives as a function of window size.}
  \label{fig:wsTP}
\end{minipage}
\hspace{0.01\linewidth}
\begin{minipage}{0.30\linewidth}
    \includegraphics[width=\linewidth]{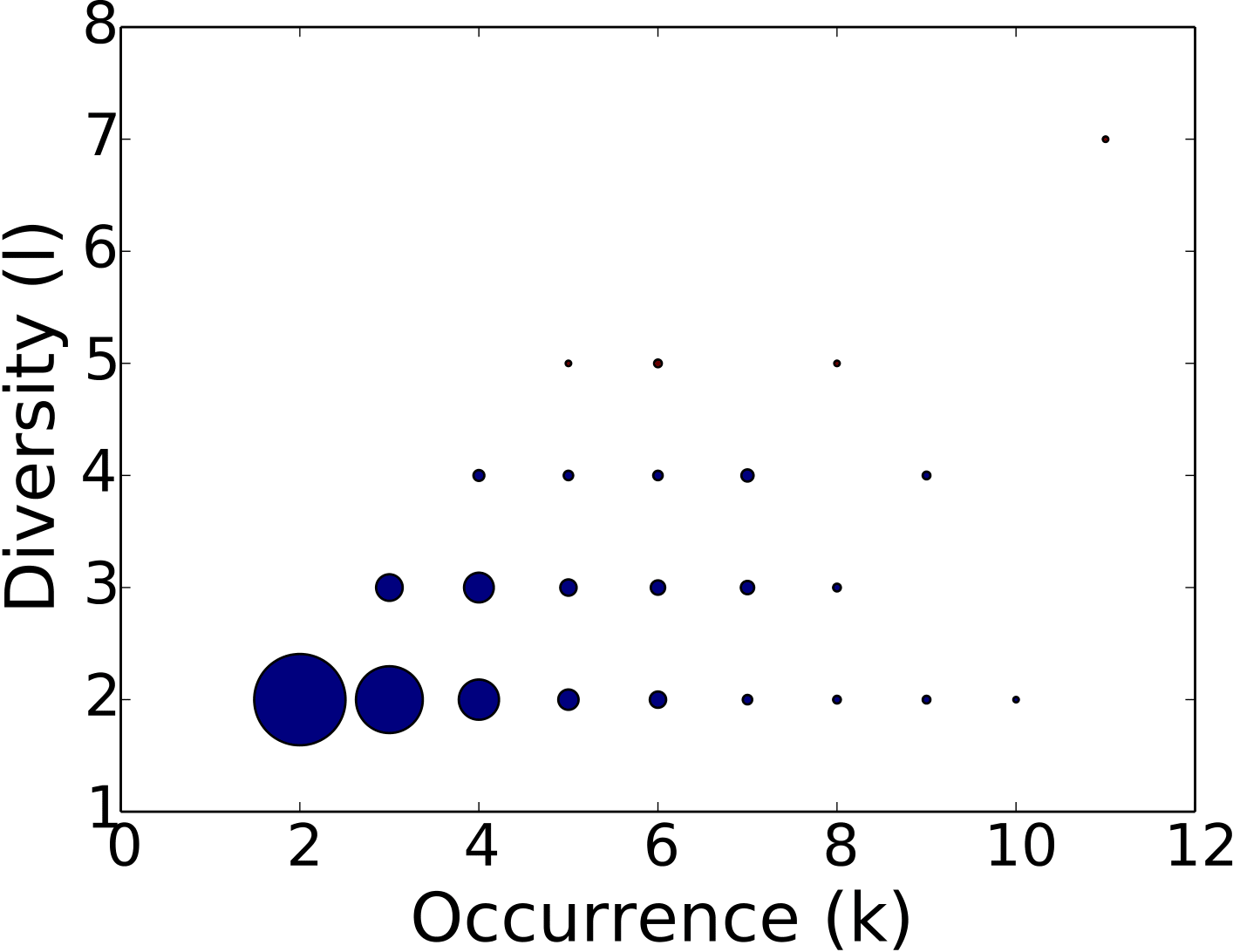}
    \caption{$k$-$l$ values distribution\hfill\eject$\;$}
    \label{fig:distribution}
\end{minipage}
\end{figure*}

\noindent We observe the running time, the number of candidate pairs, and the number of
event comparisons as a function of the dataset size. The dataset size is
increased by increasing the number of days of data included in the linkage
analysis. Furthermore, for these experiments we also change the window size.
Recall that the window size is used during the temporal filtering step to
locate co-occurring events. 

Figure~\ref{fig:performance} presents our running time related results. In all
the figures, the $x$-axis represents the dataset size in days and the $y$-axis
represents a performance metric. Different series represent varying window
sizes. 

One of the main challenges is the scalability of the linkage solution.
Processing many days of data should complete in reasonable amount of time for
the resulting analysis to be valuable. Figure~\ref{fig:runtime} plots the
running time as a function of the dataset size. We make two observations from
the figure. First, the running time of \emph{ST-Link} is
linear in the dataset size. For 5 days of data, the running time is around $1$
hour and for $40$ days of data it is around $7$ hours, all for $30$ minute
windows. Second, the running time increases with increasing window size, yet
the running time is linear in the dataset size for all window sizes.

Figure~\ref{fig:comparisons} plots the number of event-to-event comparisons as
a function of the dataset size. In our experimental evaluation, every time we
compare two location based events for either co-location or alibi check, the
number of event comparisons is increased by one. We observe that up to 15 days
of data, the number of comparisons grows at an increasing rate. Yet, after 15
days the rate starts to go down and eventually the growth of the number of
operations happens at a relatively low fixed rate. This can be explained by
the alibi checks performed by \emph{ST-Link}. Recall that when a user pair is
marked as an alibi, their records are not compared with each other anymore.
Also, if two users are marked as a candidate pair, their future records are
only compared to see if they are an alibi or not. Considering this, we can say
that within 15 days most of the candidate pairs and alibi pairs are
identified. As an important difference from the running time experiment, the
gaps between the series corresponding to the three window sizes are
considerably larger. This is because larger windows require more event to
event comparisons. Since event comparisons are not necessarily the only cost
of the algorithm (there is I/O, window processing, window index maintenance,
etc.), the running time experiment has narrower gaps between the running times
for different window sizes. The impact of these extra costs can be seen in
Figure~\ref{fig:runtime} as well; although the number of comparisons stabilize
after 15 days, the linear increase in the runtime continues.

Figure~\ref{fig:pairs} plots the number of candidate user pairs as a function
of the dataset size. Just like for the number of comparisons experiment, up to
15 days, the number of candidate pairs grows with an increasing rate and after
15 days the rate starts to decrease and eventually stabilizes at a low value.
For the case of candidate pairs, the eventual rate of increase is very low,
suggesting that observing additional data brings diminishing returns in terms
of being able to find new candidate pairs. However, this does not imply that
we are unable to perform additional linkages, because the number of linked
pairs within the candidate set can still grow (we will observe such growth
in the quality experiments).

Figure~\ref{fig:efficiency} shows the number of candidate user pairs after
each filtering step. It illustrates the effectiveness of the spatial and
temporal filtering steps of \emph{ST-Link}. If no filtering was applied on the
data, every user pair from the two datasets would have constituted a candidate
user pair. By applying only spatial filtering, the number of candidate user
pairs decreases by $43$ times compared to the no filtering case. It is
possible to say that spatial filtering is an effective step. Intuitively, if
data was spread over a wider geographical area, this step would be have been
even more effective (our datasets are limited to the geographic area of
Country X). After applying temporal filtering, the candidate user set
decreases by an additional $1,836$ times after spatial filtering.
Cumulatively, the number of candidate user pairs without any filtering is
$78,948$ times of the number of pruned candidate user pairs, which gets close
to $5$ orders of magnitude reduction in the number of pairs.

\subsection{Quality of Linkage}
\noindent We observe the precision and the number of true positives as a
function of the usage ratio, check-in probability, and window size. We also
observe the precision and recall values for a variation of the \emph{ST-Link}
algorithm that does not use weights, thus trades off precision for better
recall.

\subsubsection{Impact of Check-in Probability}
\noindent Figures~\ref{fig:meanPrecision}~and~\ref{fig:meanTP} plot the
precision and number of true positives, respectively, for the results produced
by the \emph{ST-Link} algorithm as a function of the mean check-in probability.
Different series in the figure represent different $k$-$l$ settings. We set
$k\geq l$, as the co-occurrence counts has to be greater than the diversity
counts.  For these experiments the usage ratio is set to $50\%$, which means
only half of the call users are performing check-ins.

Figure~\ref{fig:meanPrecision} shows that precision is very close to $1$ for
all $k$-$l$ settings but for $1$-$1$.  We see that using $1$-$1$ diversity
results in very poor precision for low values of the check-in probability. As
the check-in probability increases, then the precision of $1$-$1$ diversity
increases as well, but never reaches $1$. The increase is understandable, as
more events on the check-in side will help rule out incorrect candidate pairs
via alibis. Surprisingly, the precision for higher $k$-$l$ values are all
close to $1$. This is due the impact of alibi detection, and strong weight
constraint. As we will see shortly, not using weights trades off precision
for better recall. Even if two users have events that are only $2$-$2$
diverse, they can be correctly linked if they have no alibis.

Figure~\ref{fig:meanTP} shows that the number of true positives in the linkage
increases with the increasing check-in probability. This is expected, as more
events help in increasing the co-occurrence and diversity counts. We also
observe that higher $k$-$l$ values result in reduced number of linkages. Given
that $2$-$2$ diversity has very good precision, and has the second highest
true positive count (after $1$-$1$ diversity, which has unacceptable
precision), it can be considered a good setting for getting the best out of
the linkage. We see that for a check-in probability as low as $0.01$, it can
match many hundreds of users, and for probability $0.1$, it can match up to
$10$ thousand users. As we will see shortly, these numbers can be further
increased by trading off some accuracy.

\subsubsection{Impact of Usage Ratio}
\noindent Figures~\ref{fig:activePrecision}~and~\ref{fig:activeTP} plot the precision
and the number of true positives, respectively, for the results produced by the
\emph{ST-Link} algorithm as a function of the usage ratio. Different series in
the figure represent different $k$-$l$ settings, as before. For these
experiments check-in probability is taken as $0.01$.

Figure~\ref{fig:activeTP} shows that the precision of all $k$-$l$ settings is
close to $1$ throughout the entire range of the usage ratio, except for
$1$-$1$. The $2$-$2$ setting has precision values that are slightly lower than
$1$, but not lower than $0.95$. Figure~\ref{fig:meanTP} shows the true
positive counts for the same settings. As we can see clearly from the figure,
increased usage ratio results in increased number of successful linkages.
Again, this could be attributed to increasing weights for co-occurrence and
diversity, as well as increased effectiveness of alibi detection.

Interestingly, even when only $1$ percent of the call users are synthetically set to making
check-ins, and when the check-in probability around a call is set as low as $1$ in
$100$, one can still match some users (around $10$). This could also be looked
at from a privacy standpoint. In other words, being able to perform
spatio-temporal linkage across two datasets successfully even for only $10$
users may be considered a privacy breach. We plan to investigate the privacy
protection mechanisms against this kind of linkage in our future work.

\subsubsection{Unweighted Linkage}
\noindent The results so far have demonstrated high precision, but the number of
users one could match is relatively low compared to the number of total users.
In order to show the trade-off between precision and accuracy, we have also
performed experiments where the linkage model is slightly modified to use
weights that are equal to $1$. That is, we count each event co-occurrence
between two users as $1$, without considering other possible co-occurrences
these events may have with events of other users. In other words, the weight
function from Eq.~\ref{eq:weight} is taken as $1$. As it was discussed before,
there are two different approaches for deciding the values of the $k$-$l$
parameters. The first one is deciding after multiple experimental runs, and
the second one is by detecting the \emph{elbow point} of distributions of
co-occurrence and diversity values.

In this experiment we applied both to show the effectiveness of the
\emph{elbow point} detection technique as well. According to maximum absolute
second derivative test results, the values of $k$-$l$ parameters based on
elbow detection are $3$-$3$.

Table~\ref{tbl:weightTest} shows the precision and recall results for the
unweighted linkage. The recall values here represent the fraction of users
from the check-in dataset that were successfully linked. It is important to note
that we only considered users that have enough number of events. A user is said 
to have enough number of events only when she has at least $l$ diverse events, 
for each $k$-$l$ setting.
The table shows an interesting result: With unweighted linkage we see a clear
tradeoff, where with increasing $k$-$l$ values the precision improves, but the
recall drops. With the $3$-$3$ setting, we get a precision of $0.89$ and can
link $61\%$ of the users that have enough number of events. Considering all
users from the check-in dataset this value is $23~\%$. Recall that $3$-$3$
setting was identified using the \emph{elbow point}s. Increasing the diversity
setting to $5$-$5$, one gets almost perfect accuracy ($0.99$), but the recall
drops to $50\%$ of the users. When absolute accuracy is not required, such as
for machine learning to extract overall patterns, the unweighted linkage model could be more
effective in practice.

\subsubsection{Impact of Alibi}
\begin{figure}
\begin{minipage}{0.5\linewidth}
  \begin{tabular}{|r||c|c|c|c|}\hline 
          & Runtime (m)  & Precision   & Recall & Cand. Count \\ \hline 
    1   & 30     & 0.78 & 0.68     & 1,998,491   \\ \hline 
    2   & 34     & 0.75 & 0.74     & 2,651,746 \\ \hline
    4   & 39   & 0.71 & 0.82     & 3,511,090 \\ \hline
    8   & 45   & 0.68 & 0.88     & 4,446,937 \\ \hline
    16    & 58   & 0.65 & 0.91     & 5,311,043       \\ \hline
  $\infty$  & 122    & 0.62 & 0.99     & 6,765,345    \\ \hline
  \end{tabular}
\end{minipage}
  \caption{Alibi threshold experiment results.}
  \label{tbl:alibiThreshold}
\end{figure}
Alibis are used to improve both the running time performance and the accuracy.
For these experiments the check-in probability is taken as $0.5$ and the
usage ratio is taken as $50\%$. Only one grid is considered, which contains
15,268 users and 1,956,734 events in total. As it was discussed before, a threshold value
on the number of alibi events can be used before disregarding a candidate pair. In this 
experiment we evaluate the impact of alibi in terms of performance and accuracy as a function of the alibi threshold.
\medskip\\
\noindent \textbf{Performance.} 
Table~\ref{tbl:alibiThreshold} shows the running time, precision, recall, and
the number of candidate pairs for the alibi threshold experiment. When the
threshold is set to $\infty$, effectively disabling alibi detection, we
observed that the algorithm took 122 minutes to complete. At the end of the
temporal filtering step, there were 6,765,345 possible pairs. On the other
hand, when alibi is used and the threshold is set to $1$, the running time
decreased down to 30 minutes and the number of possible pairs were 1,998,491.
Almost $70~\%$ of the possible pairs were pruned with the help of alibi
detection and further processing is avoided. When larger threshold values are
used, we observe slight increase in the running time. For the threshold values
of $2$ and $16$, the processing time is $34$ and $58$ minutes, respectively.
\medskip\\
\noindent\textbf{Accuracy.} Precision of the $k$-$l$ diversity based linkage can
be increased by setting sufficiently large $k$ and $l$ values. Larger $k$ and
$l$ values decrease the probability of different users satisfying the linkage
requirements. However when at least one of the datasets is sparse, setting
larger $k$ and $l$ values will result in low recall, as many true positive
pairs will be missed. In such datasets, alibi definition prevents many false
positive pairs that satisfy the co-occurrence and diversity requirements. Our
experiments showed that when alibi is not used (threshold value $\infty$), $99
\%$ recall can be reached, yet with $62\%$ precision. In contrast, setting
alibi threshold to $1$, increases the precision to $78\%$, with recall
decreased to $68\%$. The reason behind this decrease has to do with the lack
of precise location information in our datasets. For example, when two
temporally close events of a user are from two neighboring cell towers, their
locations end up being the centers of the cell towers, as the location
information is not sufficiently fine grained. This results in
incorrectly identifying a pair of events as alibis, as the distance between
the event locations is relatively high when considering their close
timestamps. This is when the alibi threshold becomes crucial. We observe that
the recall increases to $0.74\%$ when alibi threshold set to $2$. Increasing
threshold further increases the recall values with a cost of sacrificed
precision. For alibi threshold $4$ recall is $0.82\%$ and precision is $71\%$.

\subsubsection{Window Size}
\noindent Figures~\ref{fig:wsPrecision}~and~\ref{fig:wsTP} plot the precision and
the number of true positives, respectively, as a function of the window size.
Different series in the figure represent different $k$-$l$ settings, as
before. For these experiments the check-in probability is taken as $0.01$ and the
usage ratio is taken as $50\%$. Window sizes start from $15$ minutes and
increases up to $75$ minutes in increments of $15$ minutes.

Figure~\ref{fig:wsPrecision} shows that the precision stays at $1$ is not
effected by the window size except for lines corresponding to lower $k$-$l$
values. In particular $2$-$2$ and $3$-$2$ are impacted negatively from larger
window sizes. $1$-$1$ is not shown in this experiment, as it already has a
very low precision. Note that the window size does not impact only the size
the temporal window we slide over the events, but also the definition of
co-occurrence (recall the $\alpha$ parameter from Eq.~\ref{ref:co-time}).
Increasing the window size makes it possible to match potentially
unrelated events from different real-world users and the results reflect that.
However, due to the alibi processing, the negative impact of increasing window
size on the precision is milder that it would otherwise be.

Figure~\ref{fig:wsTP} shows that the number of true positives drops with
increasing window sizes. Again this can be attributed to the increasing number
of unrelated event matches due to the larger window. Recall that if the same
user is matched to more than one user from the other dataset, we remove such
users from the linkage results. The increased window size results in ambiguity
in the results. Assuming users $x$ and $y$ are linked for a given window size,
increasing the window size does not change the fact that $x$ and $y$ are
matched, but it may result in additional matches, such as between user $x$ and
some other user $z$, and thus eliminating the correct linkage between $x$ and
$y$ from the results.

\subsubsection{$k$-$l$ value Distribution}
\noindent Figure~\ref{fig:distribution} shows the $k$-$l$ value distribution of 
user pairs after spatial and temporal linkage. The usage ratio is taken as
$50\%$ and the check-in probability is $0.01$ for this experiment. For a given
pair in the results, we find the highest $k$-$l$ diversity values it supports
and maintain these counts. In the figure, the areas of the circles are
proportional to number of pairs with the given $k$-$l$ diversity. Since the
number of pairs for $k$-$l$ lower than $2$-$2$ is too high (and precision very
low as we have seen earlier), we do not present them in the results. As
expected, the number of linked pairs is decreasing as the $k$-$l$ values are
increasing. It is interesting to note that for extreme values such as $11$-$7$
diversity, it is still possible to find user pairs. We also observe that
increasing diversity has a higher filtering power than increasing occurrence.

\subsection{Integration with SERF}
In addition to evaluating our approach under different settings, we also
attempted to integrate our linkage model with the Stanford Entity Resolution
Framework (SERF). SERF implements the R-Swoosh~\cite{ref:swoosh} algorithm.
For this integration, users are arranged as entities and their events are
considered as attributes. Given two entities, if they have enough number of
co-occurring attributes satisfying the $k$-$l$ diversity model, they are
marked as a match.

Starting with pairwise comparison of entities, R-Swoosh algorithm gradually
decreases the number of entities by merging the matching records, and deleting
the dominated ones. While this is an effective method to decrease the number
of comparisons on match heavy datasets, for datasets that contain few matching
entities, the run-time is still $\mathcal{O}(N^2)$. Merging of two
records is valid only when there is merge associativity between records. Given
three records, $r_1$, $r_2$, and $r_3$, comparisons of  $\langle r_1, \langle
r_2, r_3\rangle\rangle$ and $\langle\langle r_1,r_2\rangle, r_3\rangle$ may
result in different linkage decisions~\cite{ref:koosh}. To alleviate this
problem, the SERF framework also implements the Koosh
algorithm~\cite{ref:koosh}. Different than the R-Swoosh algorithm, when the
Koosh algorithm finds a matching pair of entities, it does not merge them
immediately, unless confidence is above a threshold. However, defining the 
confidence to use our spatio-temporal linkage model
in SERF is not straightforward and requires further research, which we
leave as future work.

Applying Koosh algorithm without using merges is almost brute force and using
a small subset our dataset (15,268 users, 1,956,734 events, in total), 
SERF takes more than $50$ hours of processing time in the same setting. In comparison,
our algorithm finds the matching users in the same dataset in $30$ minutes.

\subsubsection*{Summary}
In this experimental study, we evaluated various aspects of the $k$-$l$
diversity based linkage model and the \emph{ST-Link} algorithm. We studied the
scalability of the algorithm and showed that it scales linearly with the
dataset size. We also studied the effectiveness of the linkage and showed that
high precisions can be achieved. Using the unweighted version of our
model, some of that precision can be traded off in order to achieve better
recall values as well.

\section{Related Work}\label{sec:related} 
\noindent \textbf{Record Linkage.}
One of the earliest appearances of the term \emph{record linkage} is by
Newcombe et al.~\cite{ref:blocking1, ref:ER}.  
Several surveys exist on the
topic~\cite{ref:survey1, ref:survey2, ref:survey3}. Most of the work in this
area focus on a single type of databases and define the linked records with
respect to a similarity metric. The input to such a record linkage algorithm
is a set of records and the output from it is a clustering of records. In
contrast, our problem involves linking users from two datasets, where each
user can have multiple spatio-temporal records. A theoretical approach, and
its validation, on linking users across domains is studied
recently~\cite{ref:riederer}. The first phase this work is computing a score
for every candidate pair. In a second phase they construct a bipartite graph
of users and reduce the problem into bipartite assignment problem. Their
experiments validate the accuracy of this two phase computation. While many
works on record linkage focus on accuracy~\cite{ref:dong, ref:citations, ref:riederer} 
and a few on scalability~\cite{ref:joint}, our work must consider both. In our case,
successful linkage does not rely solely on the similarity of records and as
such \emph{ST-Link} algorithm searches multiple diverse matches, aka $k$-$l$
diversity, and also makes sure that there are no negative matches, aka alibis.
To the best of our knowledge, this is a novel approach for record linkage,
specifically targeted at spatio-temporal datasets.
\\ 
\noindent\textbf{Temporal Record Linkage and Entity Evolution.} Temporal record linkage
differs from traditional record linkage in that it takes entity evolution into
account (e.g., a person can change her phone number). The \emph{time decay}
model captures the probability of an entity changing its attribute value
within a given time interval~\cite{ref:timedecay}. The \emph{mutation model}
learns the probability of an attribute value re-appearing over
time~\cite{ref:mutation}. The \emph{transition model} learns the probability
of complex value transitions over time~\cite{ref:transition}. Furthermore,
declarative rules can be used to link records
temporally~\cite{ref:declarative}. Transition model can also capture complex
declarative rules. Temporal record linkage algorithms are able to capture the
entity evolution and determine if an entity has changed the value of one or
more of its attributes. Our problem has some resemblance to entity evolution,
since the location attributes of the users change over time. However, this
change can be better described as entity mobility, rather than entity
evolution. Application of aforementioned models to spatio-temporal datasets
might be effective in predicting a user's next stop or calculating the
probability of whether a user will return back to a given location. Yet, they
would fell short of linking spatio-temporal records of users.
\\
\noindent \textbf{Spatial Record Linkage and Spatial Joins.} Many join and
self-join algorithms are proposed in the literature for spatial
data~\cite{ref:spatialJoin}. Sehgal et al.~\cite{ref:spatialER} proposes a
method to link the spatial records by integrating spatial and non-spatial
(e.g. location name) features. However, spatial record linkage and spatial
join algorithms are not extensible to spatio-temporal data as they are based
on intersection of minimum bounding boxes, one-sided nearest join, or string
similarity. Spatio-temporal joins are more complex with constrains on both
spatial and temporal domains~\cite{ref:cstj}. Yet our problem involves more
than spatio-temporal records, it involves matching spatio-temporal record
series from two datasets.
\\ 
\noindent \textbf{Trajectory Join.}
Bakalov et al.~\cite{ref:bakalov1} define the trajectory joins as the
identification of all pairs of similar trajectories given two datasets. They
represent an object trajectory as a sequence of symbols. Based on the symbol
similarity, they prune the pairwise trajectory comparisons. Effective
evaluation of symbol similarity is supported by a tree-like index scheme.
In~\cite{ref:cstj}, the authors extend the problem to continuous queries over
streaming spatio-temporal trajectories. An important difference between
trajectory join algorithms and our work is that trajectory similarity is not
necessarily an indication of a linkage and vice verse. If one of the datasets
is denser than the other, trajectories would be dissimilar, yet we still can
have matching user pairs based on $k$-$l$ linkage. However, some indexing
structures of trajectory join algorithms are closely related to our approach.
There are multiple indexing schemes for spatio-temporal data.
In~\cite{ref:gedik,ref:gridPart, ref:gridPart2, ref:sina} various grid based
structures are used for indexing. Our spatial filtering approach is similar in
its use of a grid-based index, but instead of associating objects with grid
cells, we associate users with grid cells based on the frequency of their
events residing in these cells. There are also tree-like spatio-temporal
indexing structures, surveyed in~\cite{ref:tree-like}. A common theme of these
works is the reduction of the update cost, which is not a concern in our
work.
\\ 
\noindent \textbf{User Identification.} Our work has
commonalities with the work done in the area of user identification. For
instance, de~Montjoye et al.~\cite{ref:4point} has shown that, given a spatio-
temporal dataset of call detail records, one can uniquely identify the 95~\%
of the population by using $4$ randomly selected spatio-temporal points. Similar to our
discussion, the authors mention that  spatio-temporal points do not contribute
to information gain equally. In our work, we cover this by introducing a
\emph{weight} function. Unlike our work, \cite{ref:4point} does not consider
the linkage problem, 
instead, they study how users can be uniquely identified within a single
dataset using a small subset of their records. Another related work
is~\cite{ref:ccard}, in which authors show that using the credit card
metadata, they can identify unique users and group the transactions with
respect to users. In addition to spatio-temporal reference data, they use the
transaction price and gender as auxiliary information. Another related work is 
from Rossi et al.~\cite{ref:st-identity}, in which user identification 
techniques for GPS mobility data is presented. They
use a classification based algorithm rather than pairwise comparison of
records. Importantly, our
algorithm does not use any auxiliary information but only spatio-temporal
data, and it aims to match entities across datasets.


\section{Conclusion}\label{sec:conclusion}
\noindent In this paper, we studied the problem of matching  
real-world entities using spatial-temporal usage records from two different
LESs. By introducing the $k$-$l$--diversity model, a
novel concept that captures both spatial and temporal diversity aspects of the
linkage, we study the challenge of defining similarity between usage records
of entities from different datasets. As part of this model, we introduced
the concept of an \emph{alibi}, which effectively filters out negative
matches and significantly improves the linkage quality.

To realize the $k$-$l$--diversity model, we developed the scalable
\emph{ST-Link} algorithm that makes use of effective filtering steps.
Taking advantage of the spatial nature of the data, users are associated with
\emph{dominating grids} --- grids that contain most activities of their entities.
This enables processing each grid independently, improving scalability. Taking
advantage of the temporal nature of the data, we slide a window over both
datasets jointly and maintain set of candidate users that have co-occurring
events but no alibis. The set of candidate entities are pruned as the window is
slided.

Our experimental evaluation, conducted with several data sets showed that the running
time of the \emph{ST-Link} algorithm scales linearly with the dataset size.
Moreover, precision of the linkage results is practically $1$ for most $k$-$l$
settings. We also observed that using an unweighted version of our linkage
model, the precision can be sacrificed to achieve higher recall values. 

Spatio-temporal linkage can enable gathering large dynamic data sets for many
social good applications, such as smart cities and environmental monitoring.
We are also investigating privacy preserving methodologies, those are needed
to prevent information leakage while analyzing and sharing location based
information~\cite{ref:gedikPrivacy, ref:emreTDSC2017}.

  \bibliographystyle{abbrv} 
  \bibliography{paper}

\begin{thebibliography}{10}

\bibitem{ref:bakalov1}
P.~Bakalov, M.~Hadjieleftheriou, E.~Keogh, and V.~J. Tsotras.
\newblock Efficient trajectory joins using symbolic representations.
\newblock In {\em Int. Conf. on Mobile Data Management (MDM)}, pages 86--93,
  2005.

\bibitem{ref:cstj}
P.~Bakalov and V.~Tsotras.
\newblock Continuous spatiotemporal trajectory joins.
\newblock In {\em GeoSensor Networks}, volume 4540 of {\em Lecture Notes in
  Computer Science}, pages 109--128. Springer Berlin Heidelberg, 2008.

\bibitem{ref:koosh}
O.~Benjelloun, H.~Garcia-Molina, H.~Kawai, T.~E. Larson, D.~Menestrina, Q.~Su,
  S.~Thavisomboon, and J.~Widom.
\newblock Generic entity resolution in the serf project.
\newblock Technical Report 2006-14, June 2006.

\bibitem{ref:swoosh}
O.~Benjelloun, H.~Garcia-Molina, D.~Menestrina, Q.~Su, S.~E. Whang, and
  J.~Widom.
\newblock Swoosh: A generic approach to entity resolution.
\newblock {\em The VLDB Journal}, 18(1):255--276, Jan. 2009.

\bibitem{ref:citations}
I.~Bhattacharya and L.~Getoor.
\newblock Collective entity resolution in relational data.
\newblock {\em ACM TKDD}, 1(1):5, 2007.

\bibitem{ref:declarative}
D.~Burdick, M.~A. Hern{\'a}ndez, H.~Ho, G.~Koutrika, R.~Krishnamurthy, L.~Popa,
  I.~Stanoi, S.~Vaithyanathan, and S.~R. Das.
\newblock Extracting, linking and integrating data from public sources: A
  financial case study.
\newblock {\em IEEE Data Eng. Bull.}, 34(3):60--67, 2011.

\bibitem{ref:mutation}
Y.-H. Chiang, A.~Doan, and J.~F. Naughton.
\newblock Modeling entity evolution for temporal record matching.
\newblock In {\em ACM Int. Conf. on Man. of Data}, pages 1175--1186, 2014.

\bibitem{ref:survey1}
P.~Christen.
\newblock {\em Data matching: concepts and techniques for record linkage,
  entity resolution, and duplicate detection}.
\newblock Springer Science \& Business Media, 2012.

\bibitem{ref:index2}
P.~Christen and R.~Gayler.
\newblock Towards scalable real-time entity resolution using a similarity-aware
  inverted index approach.
\newblock In {\em Proc. of the 7th Australasian Data Mining Conf. - Volume 87},
  pages 51--60. Australian Computer Society, Inc., 2008.

\bibitem{ref:4point}
Y.-A. de~Montjoye, C.~A. Hidalgo, M.~Verleysen, and V.~D. Blondel.
\newblock Unique in the crowd: The privacy bounds of human mobility.
\newblock {\em Scientific reports}, 3, 2013.

\bibitem{ref:ccard}
Y.-A. de~Montjoye, L.~Radaelli, V.~K. Singh, et~al.
\newblock Unique in the shopping mall: On the reidentifiability of credit card
  metadata.
\newblock {\em Science}, 347(6221):536--539, 2015.

\bibitem{ref:dong}
X.~Dong, A.~Halevy, and J.~Madhavan.
\newblock Reference reconciliation in complex information spaces.
\newblock In {\em ACM Int. Conf. on Man. of Data}, pages 85--96. ACM, 2005.

\bibitem{ref:survey2}
A.~K. Elmagarmid, P.~G. Ipeirotis, and V.~S. Verykios.
\newblock Duplicate record detection: A survey.
\newblock {\em IEEE TKDE}, 19(1):1--16, 2007.

\bibitem{ref:quadTree}
R.~Finkel and J.~Bentley.
\newblock Quad trees a data structure for retrieval on composite keys.
\newblock {\em Acta Informatica}, 4(1):1--9, 1974.

\bibitem{ref:gedik}
B.~Gedik and L.~Liu.
\newblock Mobieyes: Distributed processing of continuously moving queries on
  moving objects in a mobile system.
\newblock In {\em Advances in Database Technology}, volume 2992, pages 67--87.
  Springer Berlin Heidelberg, 2004.

\bibitem{ref:gedikPrivacy}
B.~Gedik and L.~Liu.
\newblock Protecting location privacy with personalized k-anonymity:
  Architecture and algorithms.
\newblock {\em IEEE Trans. on Mobile Computing}, 7(1):1--18, 2008.

\bibitem{ref:tutorial}
L.~Getoor and A.~Machanavajjhala.
\newblock Entity resolution: Theory, practice \& open challenges.
\newblock In {\em VLDB Conf.}, 2012.

\bibitem{ref:leveldb}
S.~Ghemawat and J.~Dean.
\newblock {LevelDB}.
\newblock \url{https://github.com/google/leveldb}, 2015.

\bibitem{ref:index1}
A.~Gionis, P.~Indyk, and R.~Motwani.
\newblock Similarity search in high dimensions via hashing.
\newblock In {\em VLDB Conf.}, pages 518--529. Morgan Kaufmann Publishers Inc.,
  1999.

\bibitem{ref:blocking2}
M.~A. Hern\'{a}ndez and S.~J. Stolfo.
\newblock The merge/purge problem for large databases.
\newblock In {\em ACM Int. Conf. on Man. of Data}, pages 127--138. ACM, 1995.

\bibitem{ref:spatialJoin}
E.~H. Jacox and H.~Samet.
\newblock Spatial join techniques.
\newblock {\em ACM Trans. Database Syst.}, 32(1), Mar. 2007.

\bibitem{ref:survey3}
H.~K\"{o}pcke, A.~Thor, and E.~Rahm.
\newblock Evaluation of entity resolution approaches on real-world match
  problems.
\newblock {\em Proc. of the VLDB Endowment}, 3:484--493, Sept. 2010.

\bibitem{ref:transition}
F.~Li, M.~L. Lee, W.~Hsu, and W.-C. Tan.
\newblock Linking temporal records for profiling entities.
\newblock In {\em ACM Int. Conf. on Man. of Data}, pages 593--605. ACM, 2015.

\bibitem{ref:timedecay}
P.~Li, X.~Dong, A.~Maurino, and D.~Srivastava.
\newblock Linking temporal records.
\newblock {\em VLDB Conf.}, 4(11):956--967, 2011.

\bibitem{ref:tree-like}
Y.~Manolopoulos, Y.~Theodoridis, and V.~Tsotras.
\newblock Spatiotemporal access methods.
\newblock In {\em Advanced Database Indexing}, volume~17 of {\em Advances in
  Database Systems}, pages 141--166. Springer US, 2000.

\bibitem{ref:sina}
M.~F. Mokbel, X.~Xiong, and W.~G. Aref.
\newblock Sina: Scalable incremental processing of continuous queries in
  spatio-temporal databases.
\newblock In {\em ACM Int. Conf. on Man. of Data}, pages 623--634. ACM, 2004.

\bibitem{ref:gridPart}
K.~Mouratidis, D.~Papadias, and M.~Hadjieleftheriou.
\newblock Conceptual partitioning: An efficient method for continuous nearest
  neighbor monitoring.
\newblock In {\em ACM Int. Conf. on Man. of Data}, pages 634--645. ACM, 2005.

\bibitem{ref:blocking1}
H.~B. Newcombe and J.~M. Kennedy.
\newblock Record linkage: Making maximum use of the discriminating power of
  identifying information.
\newblock {\em Commun. ACM}, 5(11):563--566, Nov. 1962.

\bibitem{ref:ER}
H.~B. Newcombe, J.~M. Kennedy, S.~J. Axford, and A.~P. James.
\newblock Automatic linkage of vital records: Computers can be used to extract
  "follow-up" statistics of families from files of routine records.
\newblock {\em Science}, 130(3381):954--959, 1959.

\bibitem{ref:gridPart2}
J.~M. Patel, Y.~Chen, and V.~P. Chakka.
\newblock Stripes: An efficient index for predicted trajectories.
\newblock In {\em ACM Int. Conf. on Man. of Data}, pages 635--646. ACM, 2004.

\bibitem{ref:riederer}
C.~Riederer, Y.~Kim, A.~Chaintreau, N.~Korula, and S.~Lattanzi.
\newblock Linking users across domains with location data: Theory and
  validation.
\newblock In {\em Proc. of the Int. Conf.on WWW}, pages 707--719, 2016.

\bibitem{ref:st-identity}
L.~Rossi, J.~Walker, and M.~Musolesi.
\newblock Spatio-temporal techniques for user identification by means of {GPS}
  mobility data.
\newblock {\em CoRR}, abs/1501.06814, 2015.

\bibitem{ref:looseTree}
H.~Samet, J.~Sankaranarayanan, and M.~Auerbach.
\newblock Indexing methods for moving object databases: Games and other
  applications.
\newblock In {\em ACM Int. Conf. on Man. of Data}, pages 169--180. ACM, 2013.

\bibitem{ref:spatialER}
V.~Sehgal, L.~Getoor, and P.~D. Viechnicki.
\newblock Entity resolution in geospatial data integration.
\newblock In {\em Proc. of the 14th Annual ACM Int. Symp. on Adv. in Geographic
  Information Systems}, pages 83--90, 2006.

\bibitem{ref:jensen}
A.~Skovsgaard, D.~Sidlauskas, and C.~Jensen.
\newblock Scalable top-k spatio-temporal term querying.
\newblock In {\em IEEE Int. Conf. on Data Engineering}, pages 148--159, March
  2014.

\bibitem{ref:joint}
S.~E. Whang and H.~Garcia-Molina.
\newblock Joint entity resolution on multiple datasets.
\newblock {\em The VLDB Journal}, 22(6):773--795, 2013.

\bibitem{ref:payger}
S.~E. Whang, D.~Marmaros, and H.~Garcia-Molina.
\newblock Pay-as-you-go entity resolution.
\newblock {\em IEEE TKDE}, 25(5), 2013.

\bibitem{ref:emreTDSC2017}
E.~Yilmaz, H.~Ferhatosmanoglu, E.~Ayday, and R.~C. Aksoy.
\newblock Privacy-preserving aggregate queries for optimal location selection.
\newblock {\em IEEE Trans. on Dependable and Secure Computing}, 1(1), 2017.

\end{thebibliography}

\vspace{-1.4cm}
\begin{IEEEbiography}[{\vspace{-0.65cm}\includegraphics[height=1in,clip,keepaspectratio]{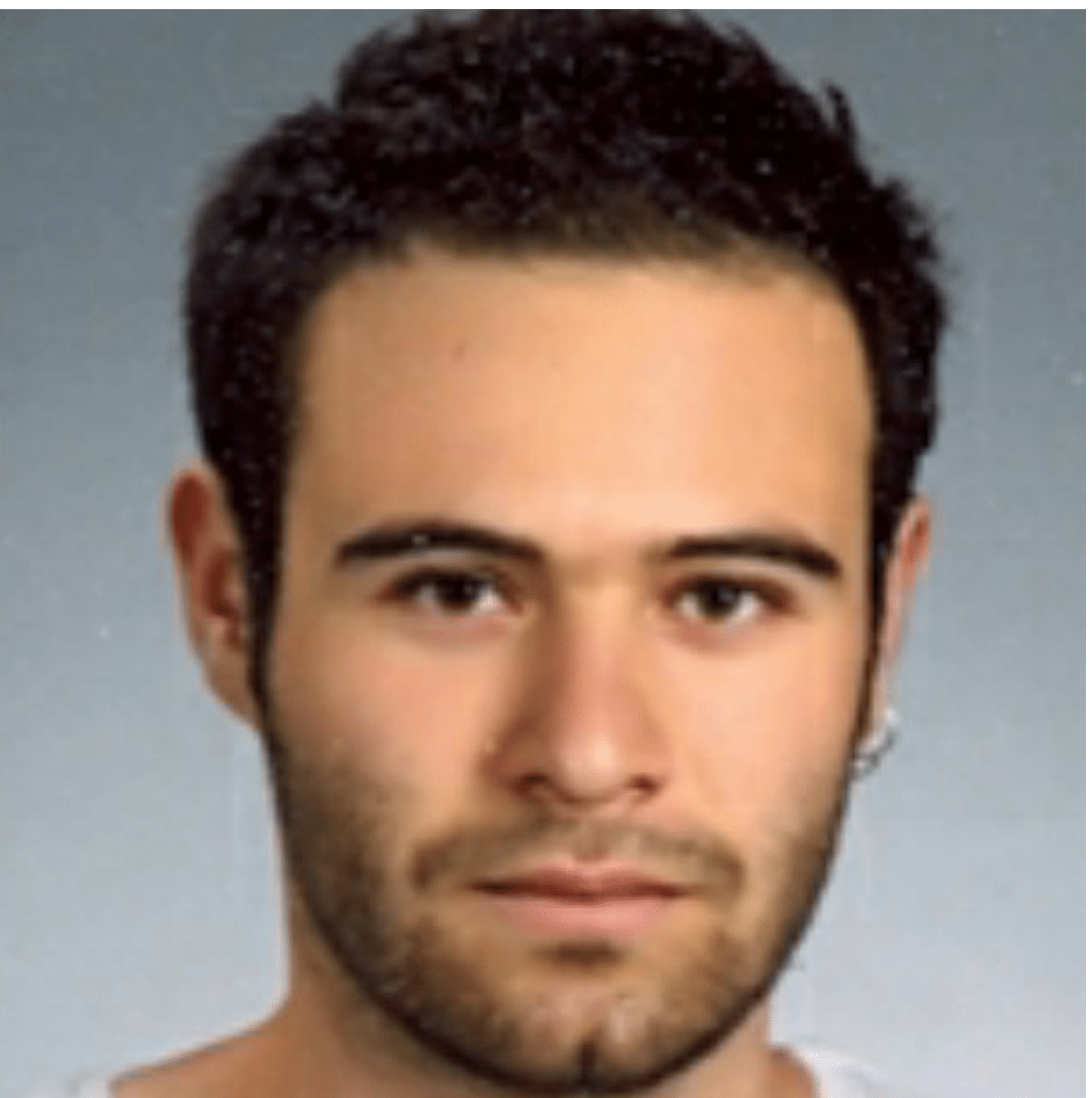}}]
{Fuat Bas{\i}k} is a graduate student in the Department of Computer
Engineering, Bilkent University, Turkey. He holds a M.Sc. degree in Computer Science from Bilkent University. His research interests are in scalable data integration.
\end{IEEEbiography}        
 \vspace{-1.8cm}
\begin{IEEEbiography}[{\vspace{-1.00cm}\includegraphics[height=1in,clip,keepaspectratio]{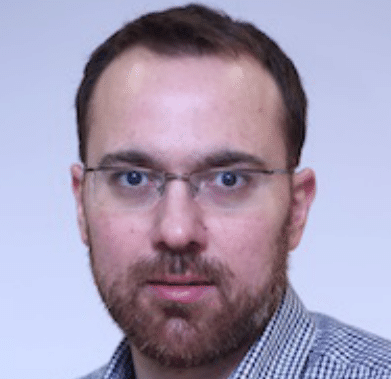}}]
{Bu\u{g}ra Gedik} is an Associate Professor in the Department of Computer Engineering,
Bilkent University, Turkey. He holds a Ph.D. degree in Computer Science from
Georgia Institute of Technology. His research interests are in data-intensive
distributed systems.
\end{IEEEbiography}         
\vspace{-1.9cm}
\begin{IEEEbiography}[{\vspace{-1.00cm}\includegraphics[height=1in,clip,keepaspectratio]{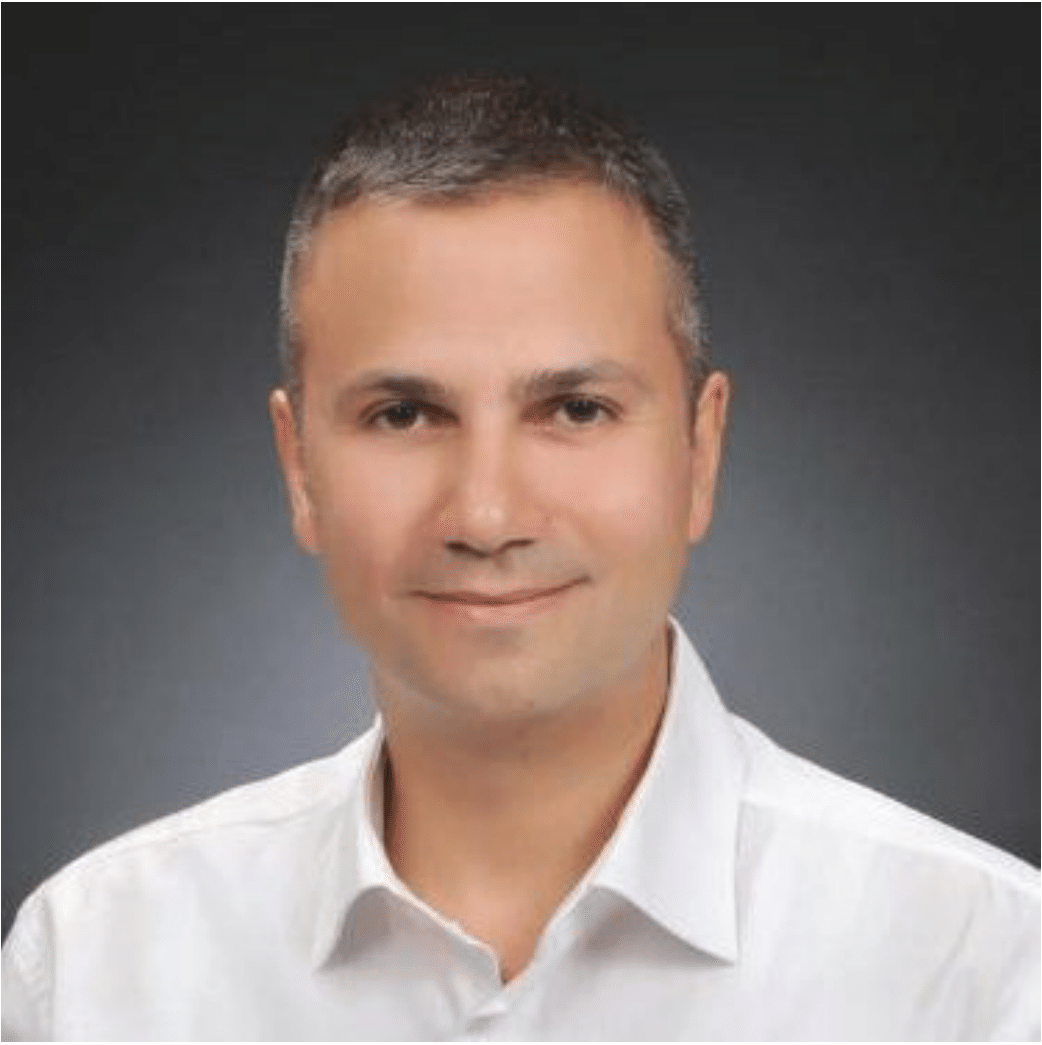}}]
{\c{C}a\u{g}r{\i} Etemo\u{g}lu} is a Manager at Turk Telekom. He holds a Ph.D. degree Electrical and
Computer Engineering from University of California, Santa Barbara. His
research interest is big data systems and applications.
\end{IEEEbiography}         
\vspace{-1.9cm}
\begin{IEEEbiography}[{\vspace{-0.65cm}\includegraphics[height=1in,clip,keepaspectratio]{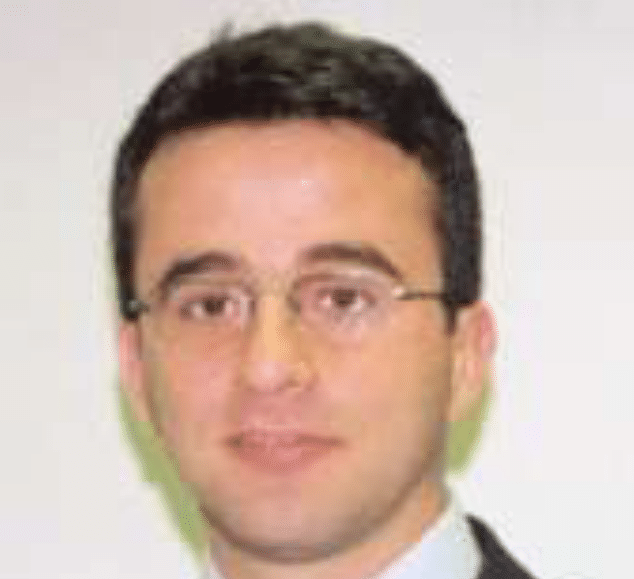}}]
{Hakan Ferhatosmano\u{g}lu} is a Professor in the Department of Science at the University of Warwick, and at Bilkent University. His research is on scalable data management and analytics for multi-dimensional data. He holds a Ph.D. degree in Computer Science from University of California, Santa Barbara. He received research career awards from the US Department of Energy, US National Science Foundation, The Science Academy of Turkey, and Alexander von Humboldt Foundation.
\end{IEEEbiography}

\end{document}